%% file: main.tex
\documentclass[pre,showpacs,notitlepage,twocolumn,superscriptaddress, amsmath,amssymb,  aps]{revtex4-2}
\usepackage{lipsum}
\usepackage{graphicx}
\usepackage{dcolumn}
\usepackage{bm}

\usepackage[dvipsnames]{xcolor}
\usepackage{tikz}

\newcommand{\revise}[1]{\textcolor{black}{#1}}

\usepackage[colorlinks=true, linkcolor=blue, citecolor=blue, urlcolor=blue]{hyperref}
\hypersetup{
    colorlinks=true,
    linkcolor=blue,
    citecolor=blue,
    urlcolor=blue
}

\newcommand{\refpanel}[2]{\hyperref[#1]{\ref{#1}(#2)}}
\newcommand{\refpanels}[3]{\hyperref[#1]{\ref{#1}(#2)--(#3)}}

\begin{document}

\title{Modulation of Non-equilibrium Structures \\ of Active Dipolar Particles by an External Field}

\author{Baptiste Parage}
 \affiliation{Institute of Theoretical Physics, University of Amsterdam, Amsterdam, The Netherlands}
 \author{Sara Jabbari-Farouji}
 \affiliation{Institute of Theoretical Physics, University of Amsterdam, Amsterdam, The Netherlands}
\date{\today}

\date{\today}

\begin{abstract}

We study the impact of an external alignment field on the structure formation and polarization behavior of low-density dipolar active particles in three dimensions. Performing extensive Brownian dynamics simulations, we characterize the interplay between long-range dipolar interactions, field alignment, and self-propulsion. We find that the competition between activity (favoring bond breaking) and the field's orientational constraint (promoting bond formation) gives rise to a rich variety of self-assembled, actuated structures. At low to intermediate field strengths, disordered fluids composed of active chains and active percolated networks can emerge, whereas strong fields drive the formation of polarized columnar clusters. Counterintuitively, low activity levels significantly extend the range of field strengths over which percolated networks persist. This structural evolution manifests in the polarization response of strongly dipolar systems, which exhibit a transition from super-Langevin to sub-Langevin behavior with increasing activity, as a result of the coupling between structure formation and activity-induced bond breaking.

\end{abstract}


\maketitle

\section{Introduction}

Over the past few decades, the engineering of microscopic self-driven, or active, particles has emerged as a rapidly expanding field, fueled by advances in the experimental realization of active colloids~\cite{Bishop} and the development of nano- and micro-robots~\cite{Zhou}. These systems are not only of fundamental interest for exploring the emergent physics of nonequilibrium systems, but are also promising candidates for the creation of a new class of active materials exhibiting novel collective properties~\cite{Bishop, Stark}. Of no less interest is the combination of self-propulsion at the single-particle level with external drives, such as magnetic or electric fields, or chemical gradients, which opens a direct route to targeted cargo delivery, with potential applications in chemistry, biology, and medicine~\cite{Qiu, Felfoul, Alsaadawi}.

Although numerous studies have explored the role of interaction type (attractive, repulsive), symmetry (isotropic, polar, nematic), and range on the collective behavior of active particles \cite{Redner2, Prymidis, Redner, Stenhammar, Meng2, Liao, Kelidou, Grossmann}, the influence of external drives on their collective dynamics remains comparatively understudied. Existing studies include the motion of active Brownian particles in gravity and harmonic traps \cite{Bickmann, Stark2}, linear concentration gradients \cite{Vinze}, external fluid flows \cite{Pedley, Meng, Stark2}, and external electric \cite{Chao} and magnetic fields \cite{Koessel, Koessel2, Telezki2, Maloney, Othman}. This latter case is of notable importance, as the orientation of the active agents can be finely steered and particles can be directed to regions of interest, provided they carry a dipole moment \cite{Alsaadawi, Baraban, Bente}. See, for instance, the recent application of magnetic dipole-bearing swimmers in external fields in targeted tumor therapy with magnetotactic bacteria~\cite{Wang2}. Moreover, for passive dipolar particles, external fields have been found to promote self-assemblies with unusual symmetries~\cite{Snezhko} and to possibly favor particle aggregation~\cite{Stevens, Jager, Jager2}, positioning them as prime tools to leverage the properties of a new class of active materials.

In the context of active dipolar particles in external fields, experimental studies on suspensions of magnetotactic bacteria~\cite{magnetotactic_bacteria_Bazylinski,MTB_24} highlighted a rich diversity of collective behaviors, including the formation of tunable three-dimensional (3D) clusters \cite{Pierce}, band- and vortex-like structures  \cite{Spormann, Guell, Vincenti}, magneto-convection patterns \cite{Thery}, and pearling instabilities \cite{Waisbord}. Furthermore, suspensions of magnetic microswimmers with hydrodynamic interactions in external alignment fields have been studied using kinetic theory~\cite{Koessel, Koessel2}, revealing emergent dynamical patterns such as sheets and columns traveling along the field direction, corroborating experimental observations. The two-dimensional (2D) self-assembly of active dipolar particles under an external alignment field has also been examined at various densities using Brownian dynamics simulations \cite{Telezki2}. The analysis showed that, while weak fields have little to no influence on the systems, strong fields prevent the formation of interconnected structures at large dipolar interactions, favoring instead ordered columnar clusters aligned with the field \cite{Telezki2}. For systems at intermediate density, the external field was found to strongly mitigate the effect of activity, in striking contrast with the zero-field case \cite{Liao}. However, how the interplay between self-propulsion, anisotropic long-range dipolar interactions, and external alignment fields impact the collective structure formation in 3D systems remains completely unexplored.  It would be interesting to see how activity and external orientational constraints affect the structure formation of dipolar particles compared to their passive counterparts \cite{DelGado, Vissers, Skipper}.

To bridge this gap and building on our previous work in the zero-field limit~\cite{Kelidou}, we here investigate the effect of a uniform magnetic field on the self-assembly of 3D active dipolar particles in the dilute regime. To this end, we perform Brownian dynamics simulations of self-propelled particles carrying permanent dipoles aligned with their propulsion velocity,  and under an external alignment field. As discussed in reference~\cite{Kelidou}, when the active force is smaller than the dipolar interaction strength at contact, active dipolar particles form isotropic fluids composed of active chains and rings, as well as active \textit{percolated networks} with enhanced bond dynamics upon an increase of dipolar coupling strength. Conversely, strong activity overrides dipolar attraction, effectively suppressing structure formation and leading to the formation of an active \textit{gas}. Here we find that, in stark contrast to the 2D case \cite{Telezki2}, weak external fields promote the formation of large-scale structures and significantly extend the stability range of active \textit{percolated networks}. At large field strengths, states consisting of polarized gases of dipolar particles and columnar clusters dominate, similar to what is observed in 2D \cite{Telezki2}. Interestingly, active dipolar systems exhibit a non-monotonic polarization response, featuring a transition from super- to sub-\textit{Langevin} behavior with increasing activity. This behavior arises from the competition between dipolar interactions, the alignment field, and activity-enhanced bond breaking, which together drive non-equilibrium  structure formation.

The remainder of this manuscript is organized as follows: In Sec.~\ref{sec:Simulation and analysis details}, we describe the simulation model, procedure, and analysis tools used for post-processing the data. We then start by presenting our results in the passive limit in Sec.~\ref{sec:passive_limit}, before highlighting the effect of activity in Sec.~\ref{sec:active_systems}. Our analysis focuses on the structural and dynamical properties of assemblies formed by active dipolar particles in external fields, including steady-state diagrams as functions of activity and field strength. Finally, we summarize our main conclusions and outline directions for future research in Sec.~\ref{sec:conclusions}.

\section{Simulation and analysis details}
\label{sec:Simulation and analysis details}
\subsection{Model system and dynamical equations}

\begin{figure}[t!]
\centering
\hspace{0cm} \includegraphics[width=8.5cm]{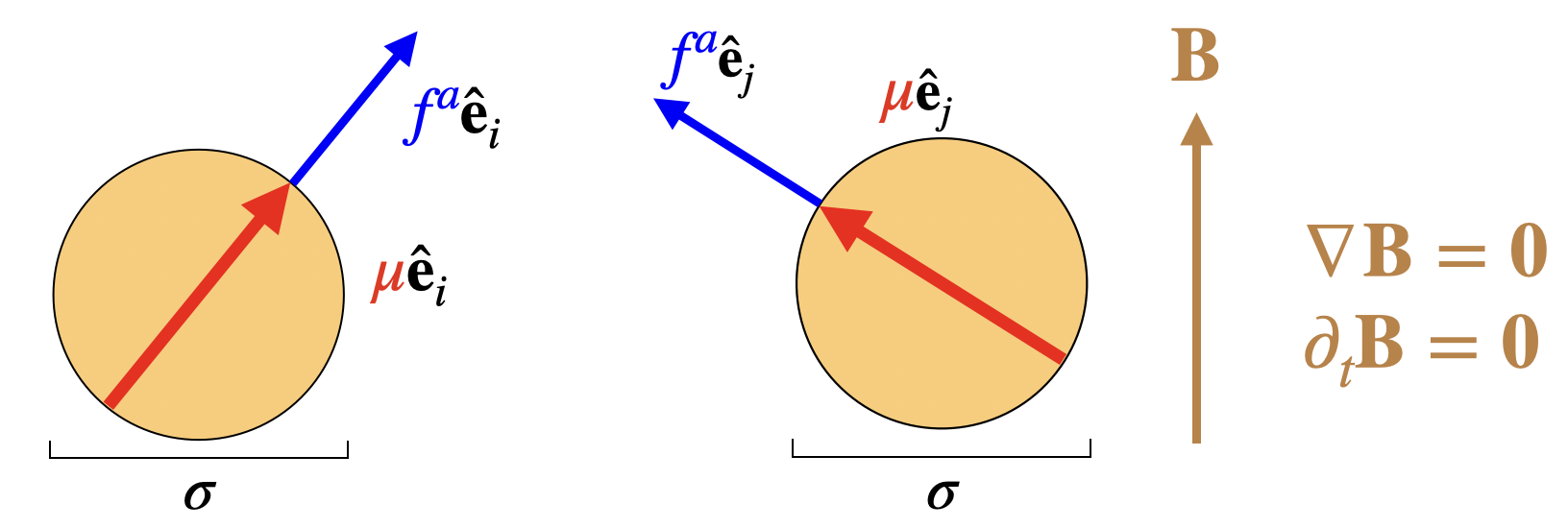}
\caption{Schematic of the model system showing two interacting spherical active dipolar particles of diameter $\sigma$. Each particle experiences a self-propulsion force $f^a \hat{\mathbf{e}}_i$ aligned with its dipole moment $\mu \hat{\mathbf{e}}_i$, and interacts with an external alignment field $\mathbf{B}$.}
\label{fig:model_system}
\end{figure}

We consider a system of $N$ active spherical particles of diameter $\sigma$, each one endowed with a point dipole moment $\bm{\mu}_i = \mu \hat{\mathbf{e}}_i$, where $i$ is the particle index, $\mu$ is the dipole's magnitude, and $\hat{\mathbf{e}}_i$ is a unit orientation vector defining its direction. Self-propulsion is modeled as for standard active Brownian particles (ABP) \cite{Callegari2019} by setting an active force $\bm{f}^a_i=f^{a} \hat{\mathbf{e}}_i$, which has a constant magnitude and is directed along the dipole moment $\bm{\mu}_i$ [Fig.~\ref{fig:model_system}]. Interparticle interactions include contributions from steric effects, modeled by the \textit{Weeks-Chandler-Anderson} (WCA)  pair potential $U_{\text{WCA}}$ \cite{WCA}, and orientation-dependent dipole-dipole interactions $U_{\text{dd}}$. The total potential energy between particles $i$ and $j$ is given by

 \begin{equation}
\label{eq:3.1}
U_{ij}(\mathbf{r}_i, \mathbf{r}_j,  \bm{\mu}_i,  \bm{\mu}_j) = U_{\text{WCA}}(r_{ij}) + U_{\text{dd}}(\mathbf{r}_{ij},  \bm{\mu}_i,  \bm{\mu}_j )
\end{equation}
\\
\noindent where $\mathbf{r}_{ij} = \mathbf{r}_{j}-\mathbf{r}_{i} = r_{ij} \hat{\mathbf{r}}_{ij}$ is the interparticle vector. The WCA potential is defined from a classic \textit{Lennard-Jones} potential, shifted by its well-depth $\epsilon$ and cut at its minimum, and reads
\begin{equation}
\label{eq:3.2}
U_{\text{WCA}}(r_{ij})=  4 \epsilon \left[\left(\frac{\sigma}{r_{ij}}\right)^{12}-\left(\frac{\sigma}{r_{ij}}\right)^{6}\right] +\epsilon,
    \end{equation}
if $r_{ij}<2^{1/6}\sigma$, and 0 otherwise. The dipole-dipole interaction potential energy is given by 
 \begin{equation}
\label{eq:u_d}
    U_{\text{dd}}(\mathbf{r}_{ij},  \bm{\mu}_i,  \bm{\mu}_j )= \frac{\mu_0 \mu^2}{4\pi} \frac{\hat{\mathbf{e}}_i \cdot \hat{\mathbf{e}}_j-3 (\hat{\mathbf{e}}_i \cdot \hat{\mathbf{r}}_{ij})(\hat{\mathbf{e}}_j \cdot \hat{\mathbf{r}}_{ij})}{r_{ij}^3}
\end{equation}
scaling with $\mu^2$ and slowly decaying as $\sim 1/r_{ij}^3$, characteristic of long-range interactions, and where $\mu_0$ is the vacuum permeability. The anisotropy of the potential $U_{\text{dd}}$ is reflected by the term $(\hat{\mathbf{e}}_i \cdot \hat{\mathbf{r}}_{ij})(\hat{\mathbf{e}}_j \cdot \hat{\mathbf{r}}_{ij})$, involving the angles between the particles’ orientations and their interconnecting vector, resulting in non-reciprocal interparticle torques. The absolute minimum of Eq.~\eqref{eq:3.1} is given by two parallel dipoles aligning "head-to-tail" ($U_{\text{dd}} \sim -2\mu^2/\sigma^{3}$), whereas a side-by-side anti-parallel configuration corresponds to a relative minimum of energy ($U_{\text{dd}} \sim -\mu^2/\sigma^{3}$). It is noteworthy that two side-by-side parallel dipoles repel each other. 

Each particle further independently interacts through its dipole with an homogeneous external field $\mathbf{B} = B\hat{\mathbf{B}}$, resulting in a dipole-field potential energy 
 \begin{equation}
\label{eq:3.4}
    U_{\text{df}} (\bm{\mu}_i, \mathbf{B})=- \bm{\mu}_i \cdot \mathbf{B} = - \mu B \hat{\mathbf{e}}_i \cdot \hat{\mathbf{B}}.
\end{equation}
We choose the field to be time-independent,  $\partial_t \mathbf{B}=\mathbf{0}$, and spatially uniform, $\nabla \mathbf{B} = \mathbf{0}$, so that no force derives from potential $ U_{\text{df}}$ and the field exerts a pure alignment torque of the form $\mu B \hat{\mathbf{e}}_i \times \hat{\mathbf{B}}$.

Particles dynamics are described by the overdamped \textit{Langevin} equations, 

\begin{align}
\dot{\mathbf{r}}_i &=   \frac{1}{\gamma_t}\left(f^{a} \hat{\mathbf{e}}_i - \sum_{j\neq i} \nabla_{\mathbf{r}_i}U_{ij} \right)
+ \sqrt{2D_t}\mathbf{\Lambda}^t_i \label{eq:3.5} \\
\dot{\hat{\mathbf{e}}}_i &=  \frac{1}{\gamma_r}\left(\sum_{j \neq i}\hat{\mathbf{e}}_i \times \nabla_{\hat{\mathbf{e}}_i}U_{ij} - \mu B \hat{\mathbf{e}}_i \times \hat{\mathbf{B}}+ \sqrt{2D_r}\mathbf{\Lambda}^r_i\right) \times \hat{\mathbf{e}}_i \label{eq:3.6}
\end{align}
\noindent where the dynamics of the active force orientation $\hat{\mathbf{e}}_i$ in Eq.~\eqref{eq:3.5} is governed by orientational fluctuations in Eq.~\eqref{eq:3.6}. Here, $D_t=k_BT/\gamma_t$ and $D_r=k_BT/\gamma_r$ represent the translational and rotational diffusion constants, respectively, where $\gamma_{t}$ and  $\gamma_{r}$ are the translational and rotational drag coefficients, and $\sigma^2 \gamma_t = 3 \gamma_r$ for spherical particles. $\boldsymbol{\Lambda}^t_i$ and $\boldsymbol{\Lambda}^r_i$
are translational and rotational white noises with zero mean and unit variance, \emph{viz.}, $\langle \boldsymbol{\Lambda}_{i}^t(t)\boldsymbol{\otimes}\boldsymbol{\Lambda}_{i}^t(t')\rangle= \textbf{1}\delta(t-t')$ and $\langle \boldsymbol{\Lambda}_{i}^r(t)\boldsymbol{\otimes}\boldsymbol{\Lambda}_{i}^r(t')\rangle= \textbf{1}\delta(t-t')$.

\subsection{Units and simulation parameters}

We choose the particle diameter $\sigma$ as the unit of length, the rotational diffusion time $\tau=1/D_r$ as the unit of time, and the thermal energy $k_B T = 1/\beta$ as the unit of energy. Following this choice, we denote all dimensionless quantities with a superscript $^*$. Notably, the rotational drag coefficient and the translational diffusion coefficient become $\gamma_r^* = 1$ and $D_t^* = D_t/(D_r \sigma^2) = 1/3$, respectively. The strength of dipole-dipole interactions, field alignment, and activity are given by the dipolar coupling strength $\lambda = \mu_0\beta \mu^2 \sigma^{-3}/(4\pi) = {\mu^*}^2$, the field coupling strength $\eta = \mu B \beta$, and the dimensionless active force $f^{a*} = \beta f^{a} \sigma$ (identical to the commonly used Péclet number \cite{Cates}), respectively.

Simulations were performed on a system of $N = 10\,648$ dipolar ABPs in a cubic box of edge size $L^*$ with periodic boundary conditions (PBC) in all three spatial directions. We made sure to avoid  significant interactions between periodic image by checking that the maximum magnitude of the dipolar potential $2\lambda/{r^*_{ij}}^3$ is much smaller than $10^{-3}$ at half-box length. A cut-off radius of $L^*/2$ was accordingly employed in the potential  of Eq.~\eqref{eq:u_d} to ensure an efficient treatment of dipolar interactions. The pairwise steric interaction strength was set to $\epsilon^*$= 1, and the particle density to $\rho^*$= 0.02 (packing fraction $\Phi \approx 0.01$) to match the dilute conditions of Ref.~\cite{Kelidou}. The dipolar coupling strength $\lambda$,  the active force $f^{a*}$ and the field coupling strength $\eta$ were varied in the ranges   $1 \le  \lambda \le 12.25$, $0 \leq f^{a*} \leq 300$, and $0 \leq \eta \leq 10$, respectively.

Brownian dynamics simulations were carried out using the LAMMPS simulation package \cite{LAMMPS} to implement and integrate equations (\ref{eq:3.5}) and (\ref{eq:3.6}), using a time-step of $dt^* = 2 \times10^{-4}$. For each investigated value of dipolar coupling strength $\lambda$, the system was first equilibrated in the passive zero-field limit, \textit{i.e.}, $f^{a*}$= 0 and $\eta = 0$,  where the initial configuration was that of randomly oriented particles located on a cubic lattice. Active systems were then equilibrated by restarting simulations from the equilibrated configuration of passive systems, setting $f^{a*}$ to the probed values while keeping $\eta = 0$. Finally, a homogeneous alignment  field was applied along the $z$-direction of the equilibrated passive and active boxes by instantaneously varying $\eta$ from 0 to the investigated values. During equilibration runs, convergence of the total potential energy, as well as polarization when $\eta \neq 0$ were monitored to ensure the systems were equilibrated. Reaching a steady state typically took from $10^2 \tau$ to $10^5 \, \tau$, depending on the $\{\lambda, f^{a*}, \eta \}$ combination. Production runs were subsequently performed on the equilibrated systems to compute the  relevant structural and dynamical properties, with typical length of $10^3 \tau$, unless otherwise stated. \revise{We verified that the WCA excluded-volume interactions remain sufficiently  hard-core even at large active forces, preventing any significant interparticle overlap, as confirmed by the radial pair distribution function (see Appendix~\ref{ap:spatial_ordering}).}



\section{Impact of external field on passive self-assembly}
\label{sec:passive_limit}

\begin{figure}[t!]
\centering
\hspace{-0.3cm} \includegraphics[width=1\linewidth]{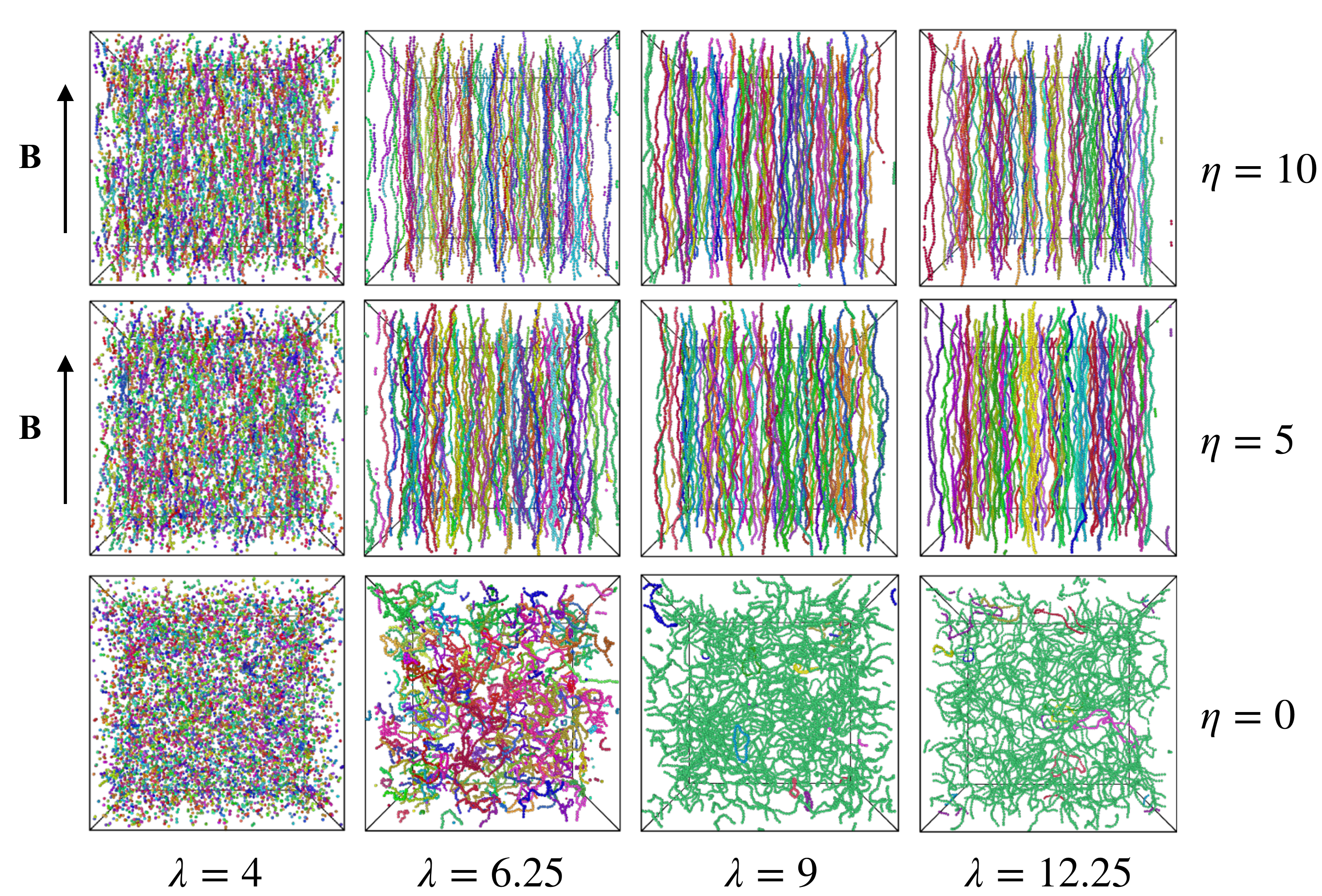}
 \caption{Representative snapshots of the passive ($f^{a*}$= 0) systems at density $\rho^*$= 0.02 for dipolar coupling strengths $\lambda$ = 4, 6.25, 9, and 12.25, and field coupling strengths $\eta$ = 0, 5, and 10. The external field $\mathbf{B}$ is indicated by the black arrows on the upper panels.}
\label{fig:passive_snapshots}
\end{figure}

\input{table_steady_states}

We start by analyzing the behavior of dipolar particles under an external field in the passive limit. Fig.~\ref{fig:passive_snapshots} presents representative snapshots of the steady-state configurations for $f^{a*}$= 0 and dipolar coupling strengths $\lambda$ = 4, 6.25, 9, and 12.25 at field strengths $\eta$ = 0, 5, and 10. In the zero-field limit (bottom panel), dipolar particles progressively self-assemble into polymeric structures as $\lambda$ increases, evolving from an apparent monomeric \textit{gas} state at $\lambda$ = 4 to finite-sized chain topologies at $\lambda$ = 6.25, before transforming into a system-spanning network at $\lambda$ = 9 and $\lambda$ = 12.25. This is the well-known percolation transition, and is consistent with previously established phase diagrams of passive dipolar colloids \cite{Stevens, Weis, Weis2, Sciortino,Kelidou}.

 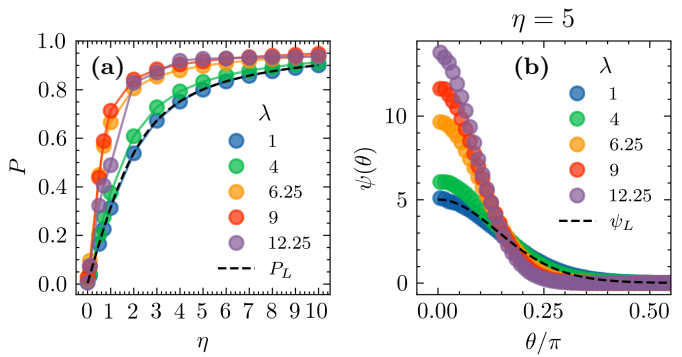
\begin{figure}[b!]
\centering
\hspace{-1cm}
\includegraphics[width=1.1\linewidth]{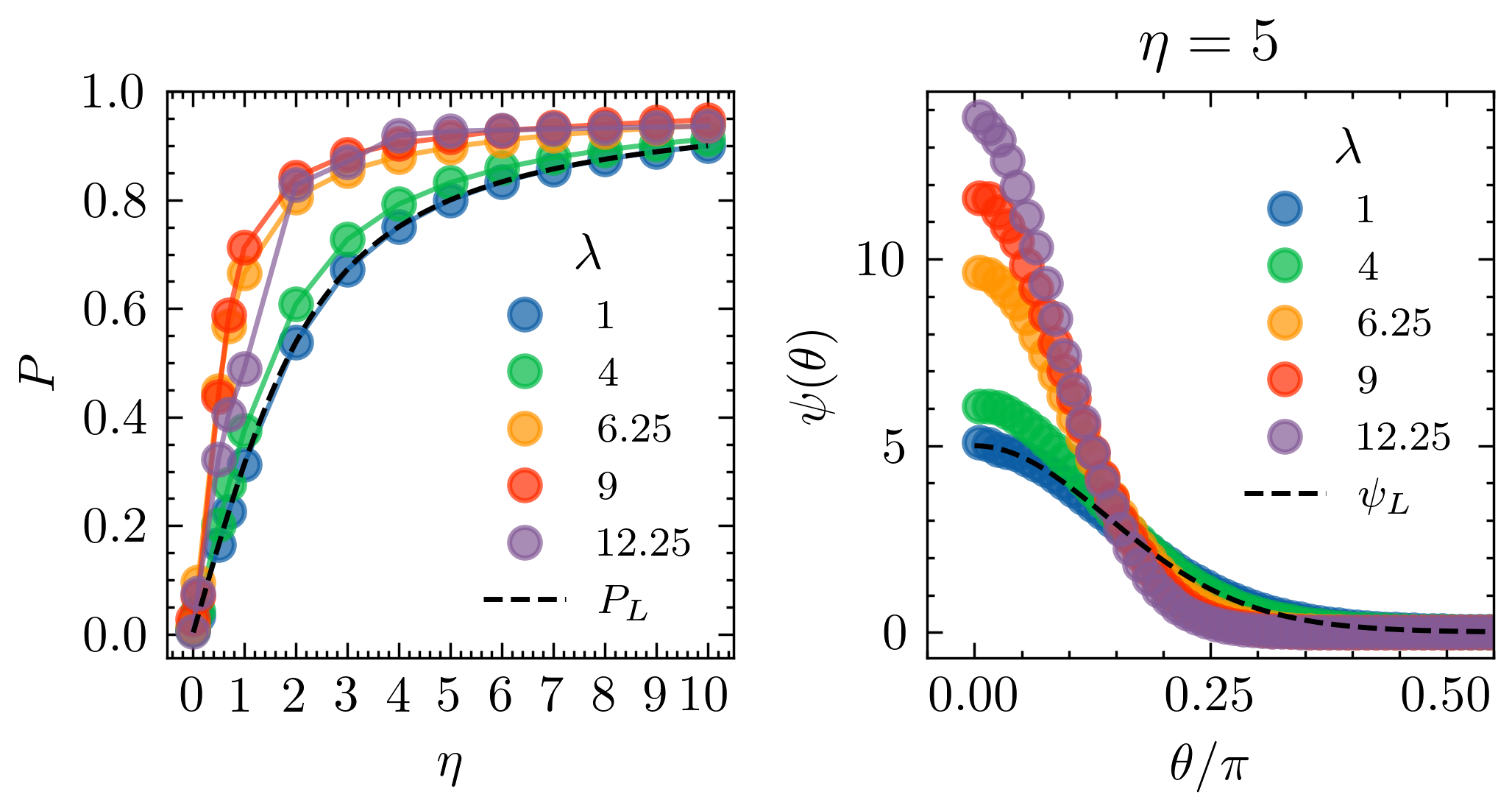}
    \caption{(\textbf{a}) Mean normalized polarization $P$ [Eq.~\eqref{eq:normalized_polarization}], as a function of the field coupling strength $\eta$ at $f^{a*}$= 0, and (\textbf{b}) angular probability distribution $\psi(\theta)$ at $f^{a*}$= 0 and $\eta = 5$, for all investigated dipolar coupling strengths $\lambda$. Dashed lines correspond to (\textbf{a}) the \textit{Langevin} function $P_L$ [Eq.\eqref{eq:langevin_polarization}], and (\textbf{b}) the \textit{Langevin} distribution $\psi_L$ [Eq.~\eqref{eq:langevin_distribution}].}
\label{fig:polarization_passive}
\begin{picture}(0,0)
  \put(-115,204){\textbf{(a)}}  
  \put(40,204){\textbf{(b)}}   
\end{picture}
\end{figure}

Applying the external field (middle and top panels), the prevailing monomeric material found at $\lambda=4$ self-assemble into small chain topologies, overall polarized in the field direction. Field-induced polymerization at this particular value of the dipolar coupling is consistent with previous simulation studies on dipolar fluids \cite{Stevens, Weis2}, and stems from the external field's stabilization of the "head-to-tail" alignment of dipoles [Sec.~\ref{subsec:polymerization_and_percolation}]. At higher dipolar coupling strengths $\lambda \geq 6.25$, the zero-field configurations, consisting either of independent chain-like topologies ($\lambda$ = 6.25) or networks ($\lambda$ = 9, 12.25), evolve into system-spanning stripe-like structures aligned with the direction of the field. These elongated assemblies are reminiscent of columnar clusters forming in ferrofluids under external magnetic fields \cite{Flores}, and are consistent with reported simulations of analogous systems \cite{Stevens, Weis2, DelGado}.

In Fig.~\refpanel{fig:polarization_passive}{a}, we present the mean normalized polarization as a function of $\eta$ for $f^{a*}$= 0 and all investigated dipolar coupling strengths, together with Eq.~(\ref{eq:langevin_polarization}) for comparison. When $\lambda$ = 1, the polarization curve matches perfectly the \textit{Langevin} function, consistent with the fact that dipole-dipole interactions are negligible at such low dipolar coupling strength and density \cite{Weis2}. Increasing $\lambda$, we find that the polarization is systematically larger than that of independent dipoles. This behavior has already been discussed in the literature~\cite{Wang, Weis2}  and was attributed to enhanced susceptibility due to the presence of polymeric self-assembled structures at high dipolar coupling strengths. As the "head-to-tail" arrangement is the preferred configuration of bonded dipoles, the alignment of one dipole with the field constrains orientations of its neighbors, resulting in an enhanced response compared to the non-interacting dipoles described by Eq.~\eqref{eq:langevin_polarization}. It is noteworthy that the low-field polarization for $\lambda$ = 12.25 is actually lower than for $\lambda$ = 6.25 and 9, as the formation of a kinetically trapped percolated network confines the system in local energy minima \cite{Sciortino}.

Fig.~\refpanel{fig:polarization_passive}{b} shows the probability density $\psi(\theta)$ of finding a dipole at a polar angle $\theta$ with respect to the field direction, for $f^{a*}$= 0 and $\eta$ = 5, at all investigated dipolar coupling strengths, along with Eq.~(\ref{eq:langevin_distribution}) for comparison. As expected, the distribution for $\lambda$ = 1 collapses onto the \textit{Langevin} distribution of non-interacting dipoles. At higher dipolar coupling strengths $\lambda \geq 4$, deviations from $\psi_L$ towards enhanced polarization become more pronounced with increasing $\lambda$, as a result of growing orientational correlation between dipoles.

\begin{figure}[b!]
\vspace{0cm}
 \includegraphics[width=1.0\linewidth]{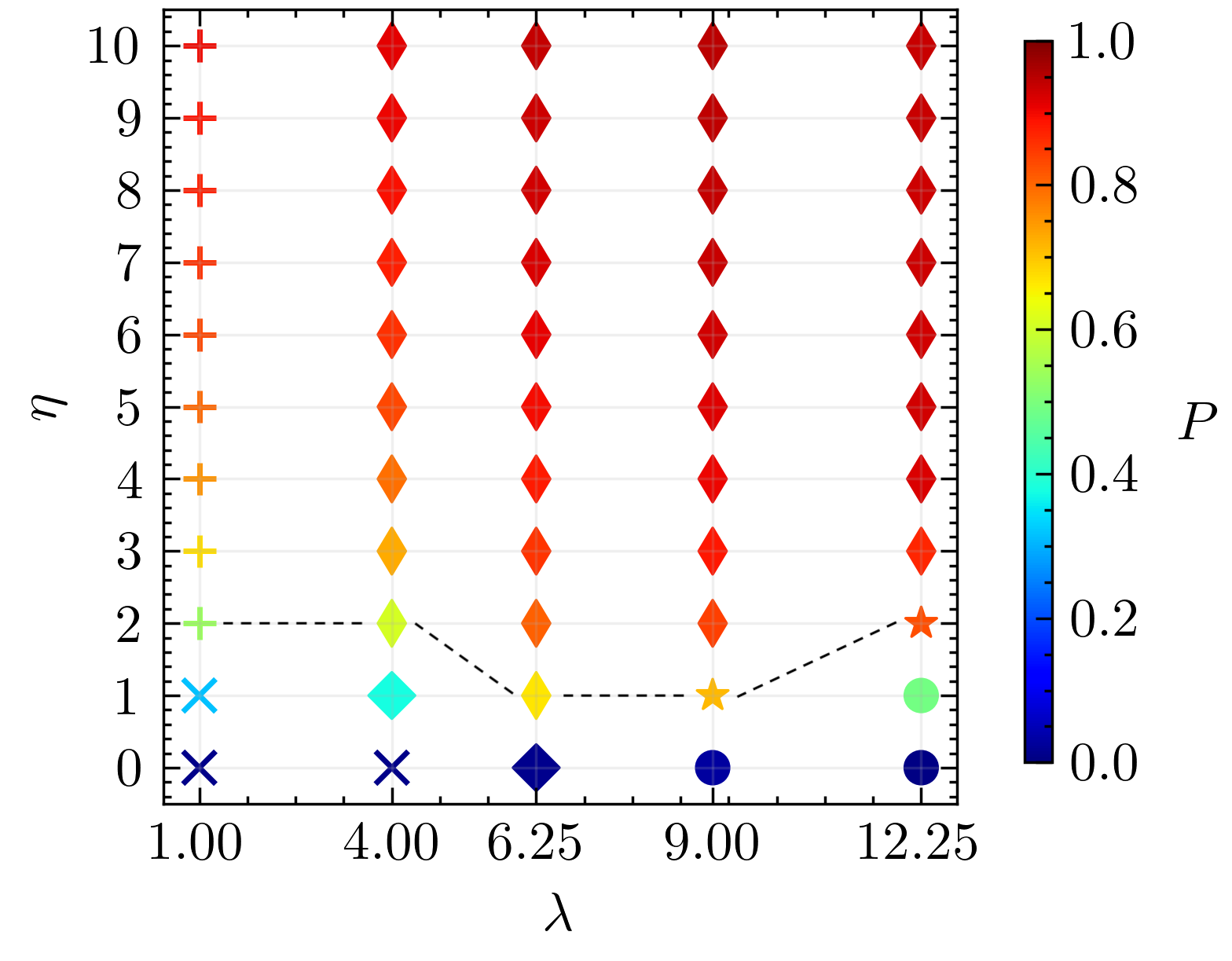}
\caption{Phase diagram of 3D passive ($f^{a*}$= 0) dipolar particles at $\rho^*$= 0.02, as a function of $\lambda$ and $\eta$. Symbols indicate different structural states based on Tab.~\ref{tab:orderparameters}: \textit{gas} (\(\bm{\times}\)), \textit{polarized gas} (\(+\)), \textit{string-fluid} (\(\diamond\)), \textit{polarized string-fluid} (\(\lozenge\)), \textit{percolated network} (\(\circ\)), and \textit{polarized \textit{percolated network}} (\(\star\)). The color bar shows the normalized polarization $P$ [Eq.~\eqref{eq:normalized_polarization}]. Dashed lines show the $P\geq0.5$ limit.}
\label{fig:phase_diagram_passive}
\end{figure}

We conclude this section by presenting the phase-diagram of passive dipolar particles as a function of $\lambda$ and $\eta$, based on the following systematic classification.
\revise{To do so, we employ a combination of three order parameters \cite{Liao, Kelidou}: the fraction of polymerized particles $\phi_p$, the fraction of particles in the largest cluster $\phi_{\text{max}}$, and the polarization $P$ (see Appendix~\ref{ap:analysis_details} for definitions). More precisely, the term \textit{gas} is employed to refer to systems where most particles are in a monomeric state, with limit fixed at $\phi_p<0.5$, further implying that $\phi_{\text{max}}<0.5$. The term \textit{string fluid} is used to refer to systems in which more than $50\%$ of the particles are in polymerized form ($\phi_p\geq0.5$) but without large-scale connectivity, the threshold being set at $\phi_{\text{max}}=0.7$, aligning with visual inspections of the simulation boxes and previous studies \cite{Sciortino, Kelidou}. Lastly, a (system-spanning) \textit{percolated network} is identified whenever $\phi_{\text{max}}\geq0.7$,  further implying $\phi_{p}\geq0.5$. To account for the effect of the external field, we differentiate between orientationally ordered and disordered states by introducing the term “Polarized” as a prefix for the latter denominations whenever $P\geq 0.5$. The classification of structural states based on order parameters $\phi_p$, $\phi_{\text{max}}$, and $P$ is summarized in Tab.~\ref{tab:classification}.}

In the limit of no activity, the field primarily induces a polarization of the pre-existing structures, with the system developing orientational order along the field direction. Thus, \textit{gases} transform into \textit{polarized gases}; \textit{string fluids} transform into \textit{polarized string fluids}; \textit{percolated networks} first progressively polarize while retaining their interconnected configurations, and ultimately transform into \textit{polarized string fluids} at stronger fields, indicating the loss of network structure in favor of independent columnar clusters, as observed in Fig.~\ref{fig:passive_snapshots}. Polarization naturally occurs earlier in systems with higher $\lambda$ values (with the exception of $\lambda = 12.25$, as discussed above), owing to cooperative dipolar interactions. Furthermore, we find that the zero-field \textit{gas} present at $\lambda = 4$ transitions into a \textit{string fluid}, and eventually into a \textit{polarized string fluid}, with increasing $\eta$. This attests to the propensity of an external field to favor polymerization, as previously noted.

\section{Impact of external field on structure formation of active dipolar particles}
\label{sec:active_systems}
\subsection{Steady-state diagrams of active dipolar particles in external field}
\label{subsec:active_systems}

\begin{figure}[b!]
  \hspace{-0.2cm} \includegraphics[width=1\linewidth]{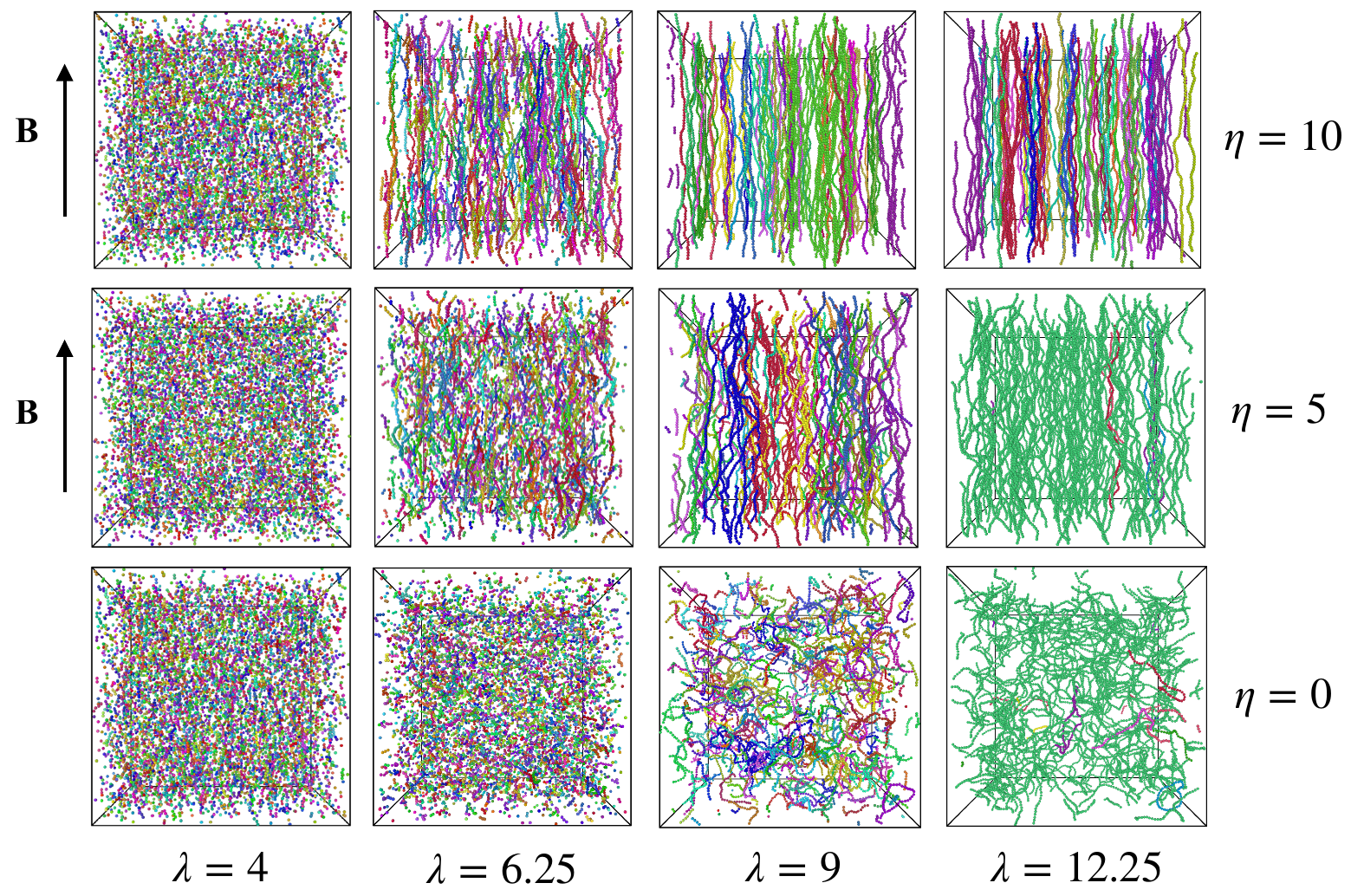}
 \caption{Representative snapshots of the active ($f^{a*}$= 20) systems at density $\rho^*$= 0.02 for dipolar coupling strengths $\lambda$ = 4, 6.25, 9, and 12.25, and field coupling strengths $\eta=$ 0, 5, and 10. The external field $\mathbf{B}$ is indicated by the black arrows on the upper panels.}
\label{fig:active_snapshots}
\end{figure}

 \begin{figure}[t!] \hspace{-1.25cm}
 \hspace{0.22cm} \includegraphics[width=9.5cm]{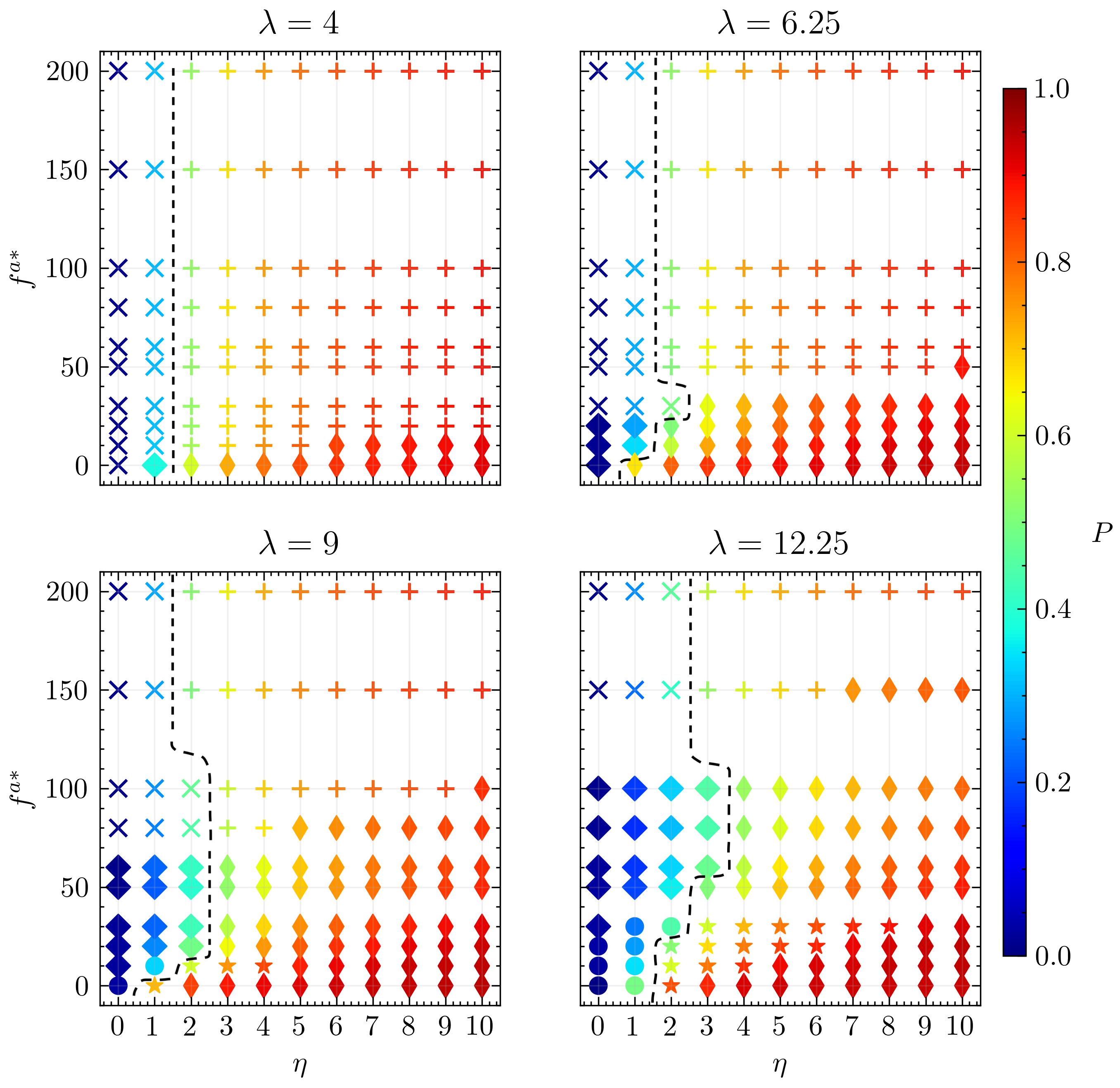}
  \caption{Steady-states diagrams of 3D dipolar ABPs at $\rho^*$= 0.02 as a function $\eta$ and $f^{a*}$ for dipolar coupling strengths $\lambda=$ (\textbf{a}) 4, (\textbf{b}) 6.25, (\textbf{c}) 9, and (\textbf{d}) 12.25. Symbols indicate different structural states based on Tab.~\ref{tab:classification} and are coded as in Fig.~\ref{fig:phase_diagram_passive}. The color bar shows the normalized polarization $P$ [Eq.~\eqref{eq:normalized_polarization}]. Dashed lines show the $P\geq0.5$ limit. The \textit{x}- and \textit{y}-axes are shared across columns and rows, respectively.}
\begin{picture}(0,0)
  \put(-120,326){\textbf{(a)}}  
  \put(-4,326){\textbf{(b)}}  
  \put(-120,200){\textbf{(c)}} 
  \put(-4,200){\textbf{(d)}}  
\end{picture}
   		\label{fig:steady_state_diagrams}
\end{figure}
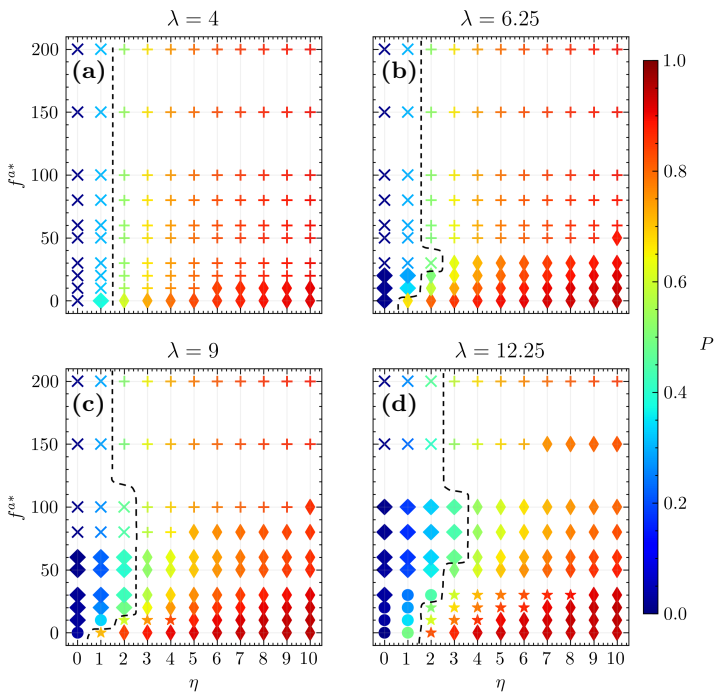

Having discussed the effect of an external field in the passive limit, we now examine its impact on the behavior of dipolar particles when activity is turned on. Fig.~\ref{fig:active_snapshots} shows representative snapshots of the steady-state configurations at $f^{a*}$= 20, for dipolar coupling strengths $\lambda =$ 4, 6.25, 9, and 12.25, and field coupling strengths $\eta=$ 0, 5 and 10. In the zero-field limit (bottom panel), similar to our previous findings \cite{Kelidou}, we generally observe the breakage of the passive self-assembled structures into smaller fragments due to activity. Passive polymeric aggregates present at $\lambda = 6.25$ are almost completely destroyed. The system-spanning network observed in the passive limit at $\lambda = 9$ breaks down into shorter chains and branched structures. At $\lambda = 12.25$ and $f^{a*}$= 20, the system retains its interconnected network but with enhanced bond breaking-reconfiguration dynamics compared to its passive counterpart [Sec.~\ref{subsec:bond_dynamics}] \cite{Kelidou}. 

Upon turning on the external field (middle and top panels), interesting structures emerge when $\lambda$ is sufficiently large. At the intermediate value of $\lambda = 6.25$, monomeric and short active polymeric structures present at zero-field self-assemble into longer chained topologies overall aligned with the direction of the field, which keep growing in size with further increasing $\eta$. At $\lambda = 9$, the long chains and branched structures found at $\eta = 0$ arrange into several elongated, branched columnar clusters spanning the whole simulation box, similar to the passive case. At the highest dipolar coupling strength of $\lambda = 12.25$, the percolated network structure,  observed in the zero-field limit, persists at intermediate field values $\eta \sim 5$, although it appears with branches aligned with the external field, before it eventually transforms into columnar structures at $\eta = 10$. The effect of activity on the spatial ordering of columnar clusters is discussed in Appendix \ref{ap:spatial_ordering}.

Figs.~\refpanels{fig:steady_state_diagrams}{a}{d} present the steady-state diagrams of dipolar ABPs as a function of $\eta$ and $f^{a*}$ for dipolar coupling strengths $\lambda = 4$, 6.25, 9, and 12.25. The classification of states follows the criteria introduced in Sec.~\ref{sec:passive_limit}. At the dipolar coupling value of $\lambda=4$, zero-field active systems remain in the \textit{gas} state. As $\eta$ increases, field-induced polymerization is generally not observed, and systems simply transition to \textit{polarized gases}. An exception is observed for $f^{a*}$= 10 where a transition toward a \textit{polarized string fluid} occurs, albeit requiring a much higher field strength than in the passive case. Furthermore, the polarization, indicated by the color overlay, appears largely independent of activity. This contrasts with higher values of the dipolar coupling ($\lambda \geq 6.25$), where the $P \geq 0.5$ threshold lines (dashed lines) exhibit a dip at intermediate activity levels, indicating that the transition to \textit{polarized} states shifts to higher field strengths compared to both low and high activity. Interestingly, this appears at values of $f^{a*}$ near the \textit{string fluid} to \textit{gas} transition, \textit{i.e.}, where the systems consists of many independent active polymerized structures. As will be discussed in Sec.~\ref{subsec:polarization}, this change actually corresponds to a transition from an enhanced to a hindered polarization response.

Focusing on the effect of the field on large-scale structures, \textit{string fluids} and \textit{percolated networks} eventually evolve into \textit{polarized} columnar structures, similar to the passive case. However, for dipolar coupling strengths $\lambda = 9$ and $\lambda = 12.25$, we observe that zero-field \textit{string fluids} formed at low activity (typically near the percolation threshold) can transition into \textit{percolated networks} at low field strengths, indicating that the combination of  sufficiently weak internal and external drives can promote connectivity in active systems. See, for example, the cases at dipolar coupling strength $\lambda = 9$ (12.25) and activity $f^{a*}$= 10 (30). Quite remarkably, active networks also exhibit greater resilience to the field than their passive counterparts, maintaining their interconnected structures up to higher field strengths. For instance, percolation at $\lambda=12.25$ breaks down above $\eta = 2$ in the passive limit, while it persists up to $\eta=$ 4 and 8 at $f^{a*}$= 10 and 30, respectively. It is noteworthy that this behavior is rather counterintuitive, for one would expect a greater susceptibility, and thus a faster destruction of the network, for a system with weaker bonds \cite{Kelidou}. However, as we will see in Sec.~\ref{subsec:polymerization_and_percolation}, the randomizing effect of activity and the field-induced polarization act cooperatively to maintain the interconnectedness of the network.

In summary, the external field modulates polarization, percolation, and polymerization in active systems, stabilizing or facilitating transitions to states with greater connectivity. In the following sections, we focus on a more detailed characterization of the underlying mechanisms by which the external field influences these newly emerging active structures.

\subsection{Field-induced polarization}
\label{subsec:polarization}

\begin{figure}[b!] 
    \centering
    \includegraphics[width=1.1\linewidth]{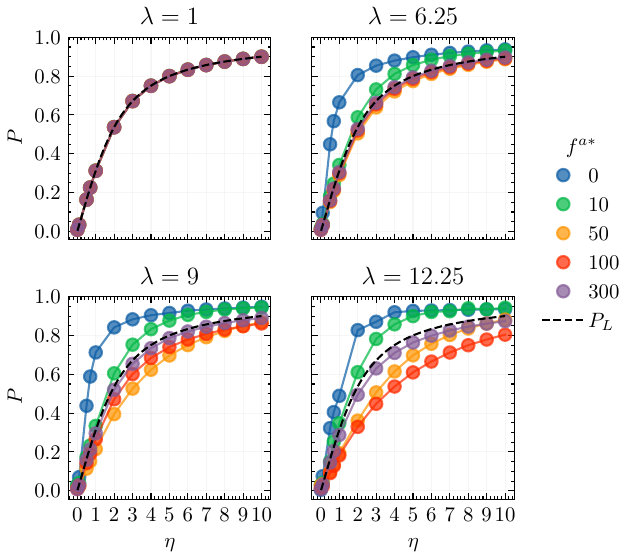}
    \caption{Mean normalized polarization $P$ [Eq.~\eqref{eq:normalized_polarization}], as a function of the field coupling strength $\eta$ at dipolar coupling strength $\lambda=$ (\textbf{a}) 1, (\textbf{b}) 6.25, (\textbf{c}) 9, and (\textbf{d}) 12.25, and for active forces $f^{a*}$= 0, 10, 50, 100, and 300. Dashed lines correspond to the \textit{Langevin} function $P_L$ [Eq.~\eqref{eq:langevin_polarization}]. The \textit{x}- and \textit{y}-axes are shared across columns and rows, respectively.}

    \begin{picture}(0,0)
  \put(-91,298){\textbf{(a)}}  
  \put(13,298){\textbf{(b)}}  
  \put(-91,187){\textbf{(c)}} 
  \put(13,187){\textbf{(d)}}  
\end{picture}
    \label{fig:polarization_active}
\end{figure}
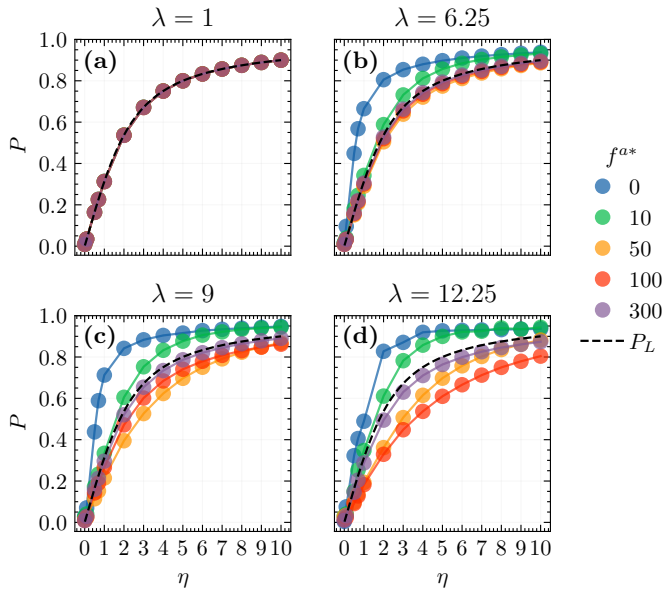

We investigate the orientational response of dipolar particles to the external field by plotting in Figs.~\refpanels{fig:polarization_active}{a}{d} the mean polarization as a function of $\eta$ for  dipolar coupling strengths $\lambda=$ 1, 6.25, 9, and 12.25, and various active forces. For comparison, we also include the \textit{Langevin} function [Eq.~\eqref{eq:langevin_polarization}], describing the polarization of non-interacting dipoles. When $\lambda = 1$, $P$ is independent of activity, with all curves matching the \textit{Langevin} function. Conversely, the polarization response is found to non-monotonically depend on the active force for higher $\lambda$ values. In the passive limit, $P$ exhibits super-\textit{Langevin} behavior originating from the presence of self-assembled structures \cite{Wang, Weis2}. In contrast, increasing activity initially leads to a decrease in polarization below the \textit{Langevin} function (sub-\textit{Langevin} behavior), followed by convergence toward it in the limit of strong activity. This super- to sub-\textit{Langevin} transition becomes more pronounced with increasing $\lambda$ and clearly indicates a transition from cooperative to inhibited dipole behavior, as previously noted in Fig.~\ref{fig:steady_state_diagrams}. Here, contrary to the passive case, the depolarization of one dipole induces the depolarization of other dipoles, leading to an overall reduction in polarization.

\begin{figure}[t!] 
   \centering
\hspace{-1cm}
\includegraphics[width=1.07\linewidth]{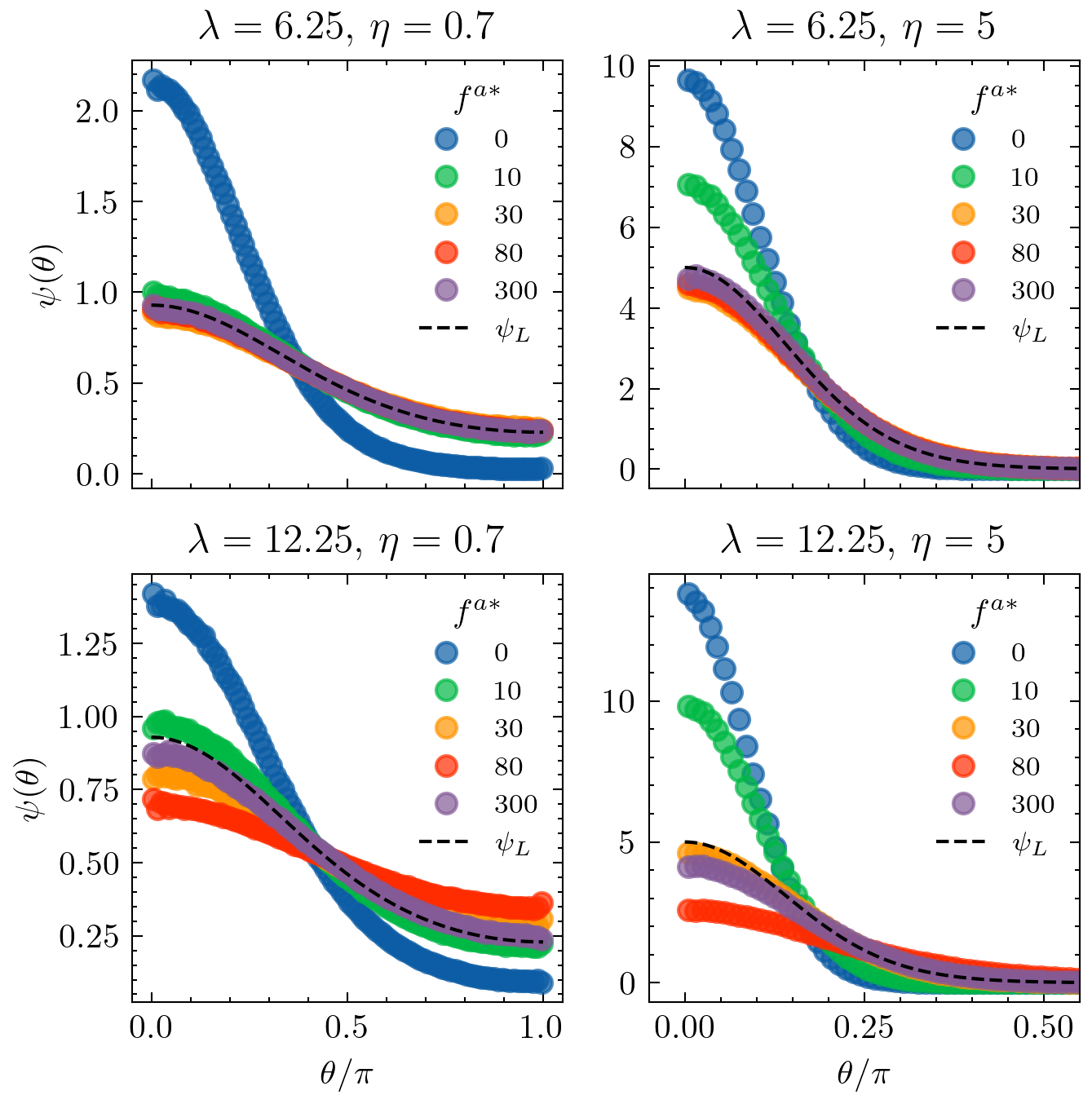}
   \caption{Angular probability distribution $\psi(\theta)$ at ($\lambda$, $\eta$) = (\textbf{a}) ($6.25$, $0.7$), (\textbf{b}) ($6.25$, $5$), (\textbf{c}) ($12.25$, $0.7$), (\textbf{d}) ($12.25$, $5$), and active forces $f^{a*}$= 0, 10, 30, 80, and 300. Dashed lines correspond to the \textit{Langevin} distribution $\psi_L$ [Eq.~\eqref{eq:langevin_distribution}]. The \textit{x}-axis is shared across columns.}
    \begin{picture}(0,0)
  \put(-77,317){\textbf{(a)}}  
  \put(48,317){\textbf{(b)}}  
  \put(-76,193){\textbf{(c)}} 
  \put(48,193){\textbf{(d)}}  
\end{picture}
   
   \label{fig:distributions_panels}
\end{figure}
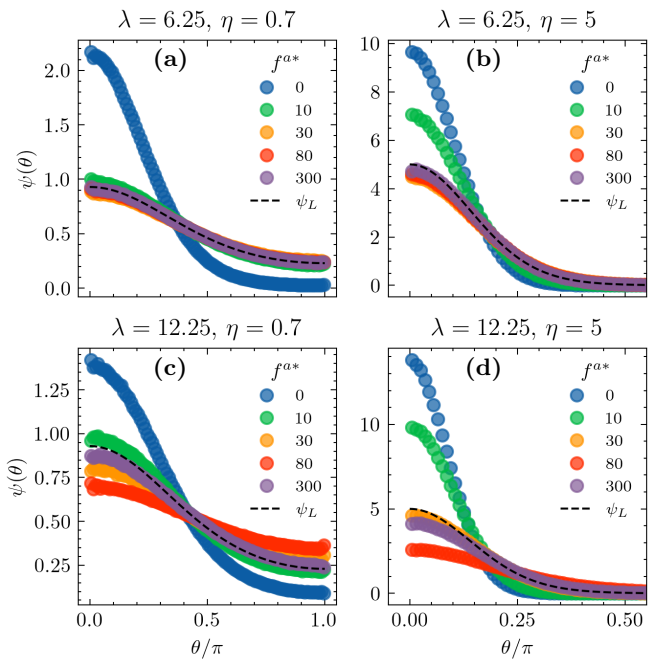

\begin{figure}[b!]
\includegraphics[width=\linewidth]{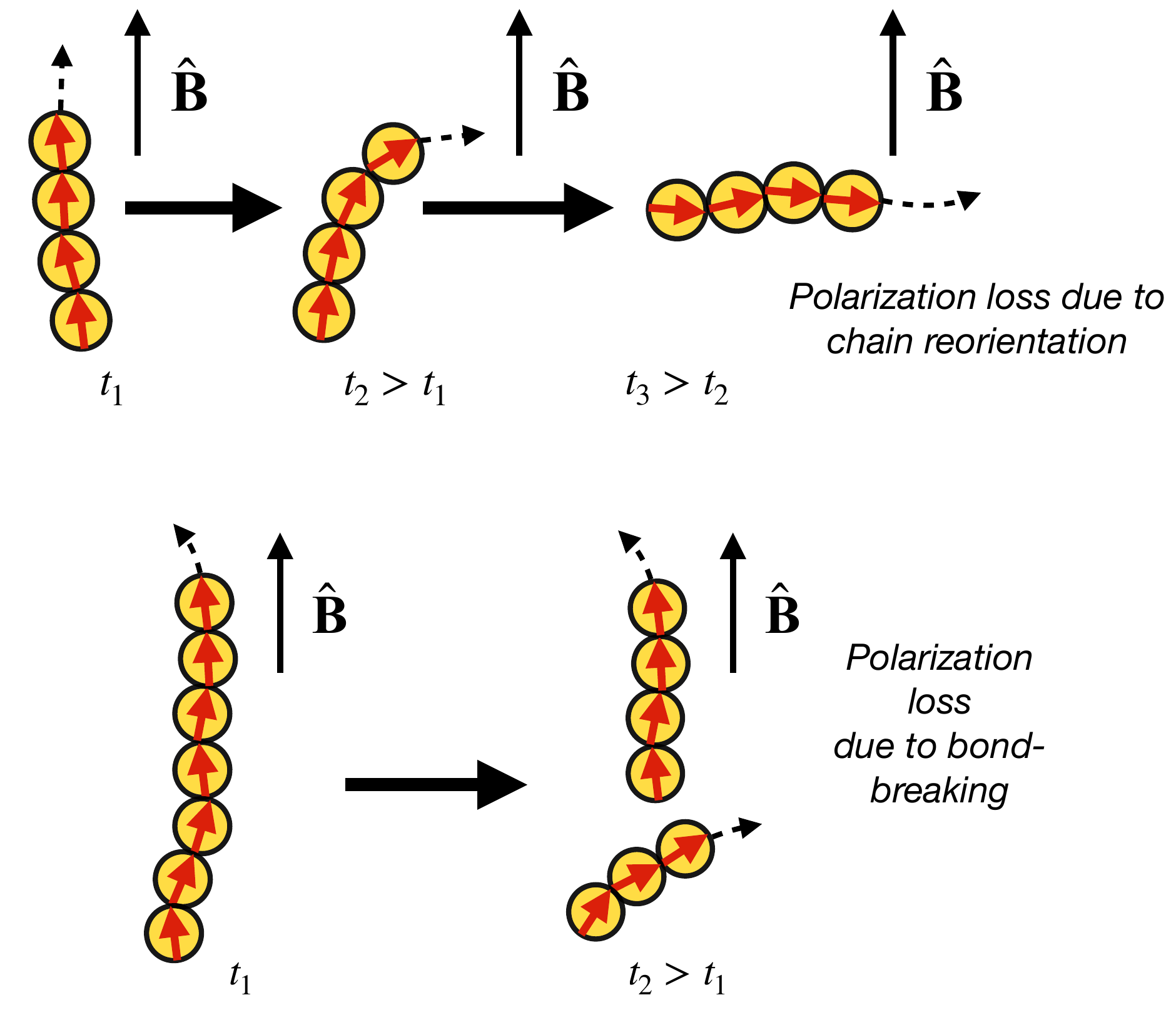}
\caption{Schematic of the polarization-loss mechanism of self-assembled dipolar ABPs \textbf{(a)} by reorientation, and \textbf{(b)} by bond-breaking.}

    \begin{picture}(0,0)
  \put(90,244){\textbf{(a)}}  
  \put(65,145){\textbf{(b)}}  
\end{picture}

\label{fig:pol_loss_schematic}
\end{figure}

To better understand the interplay between internal and external drives, we investigate the angular distribution of dipole orientations $\psi(\theta)$ relative to the external field direction. In Fig.~\ref{fig:distributions_panels}, we plot $\psi(\theta)$ at weak and intermediate field strengths $\eta = 0.7$ and $\eta = 5$, for dipolar coupling strengths $\lambda = 6.25$ (panels~(\textbf{a}) and~(\textbf{b}), respectively) and $\lambda = 12.25$ (panels~(\textbf{c}) and~(\textbf{d}), respectively), and for various active forces. As already discussed in Sec.~\ref{sec:passive_limit}, in the passive limit, the distribution for sufficiently strong dipolar coupling strengths is more peaked than the \textit{Langevin} distribution $\psi_L$ given by Eq.~\eqref{eq:langevin_distribution}. Upon increasing $f^{a*}$, the distribution progressively widens and eventually converges to $\psi_L$ in the limit of high activity. At dipolar coupling $\lambda$ = 6.25, where systems rapidly transform into a \textit{gas} with increasing $f^{a*}$, $\psi(\theta)$ converges to $\psi_L$ from above, and does so more quickly at $\eta$ = 0.7 than at $\eta$ = 5. This indicates a latency in the system's transition to a \textit{gas} state at higher field strengths, consistent with the observations made in Fig.~\ref{fig:steady_state_diagrams} that the field can favor particle aggregation. At $\lambda = 12.25$, where polymerized states persist up to intermediate activities, the width of the distribution evolves non-monotonically. The convergence of $\psi(\theta)$ to $\psi_L$ occurs in two step, with the distribution first falling below the \textit{Langevin} prediction, then converging toward it from below. This pattern is independent of $\eta$ and is consistent with the mean polarization behavior observed in Fig.~\ref{fig:polarization_active}.

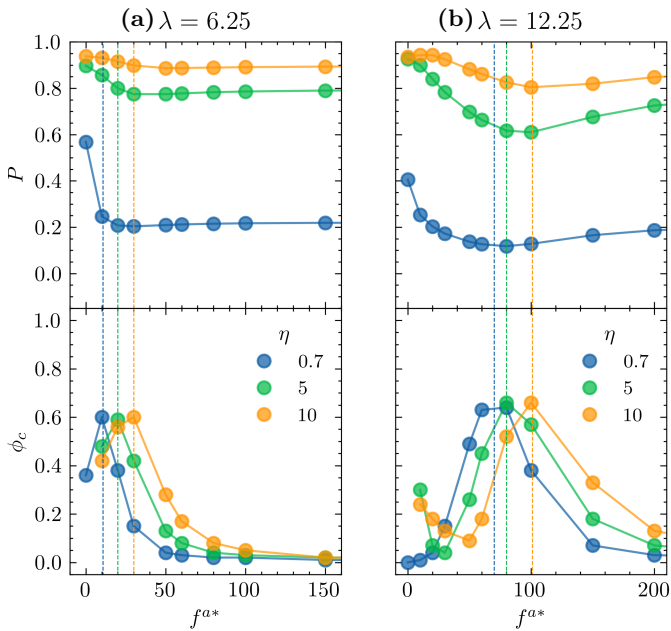
\begin{figure}[t!]
\hspace{-1cm}
\centering
\includegraphics[width=1.05\linewidth]{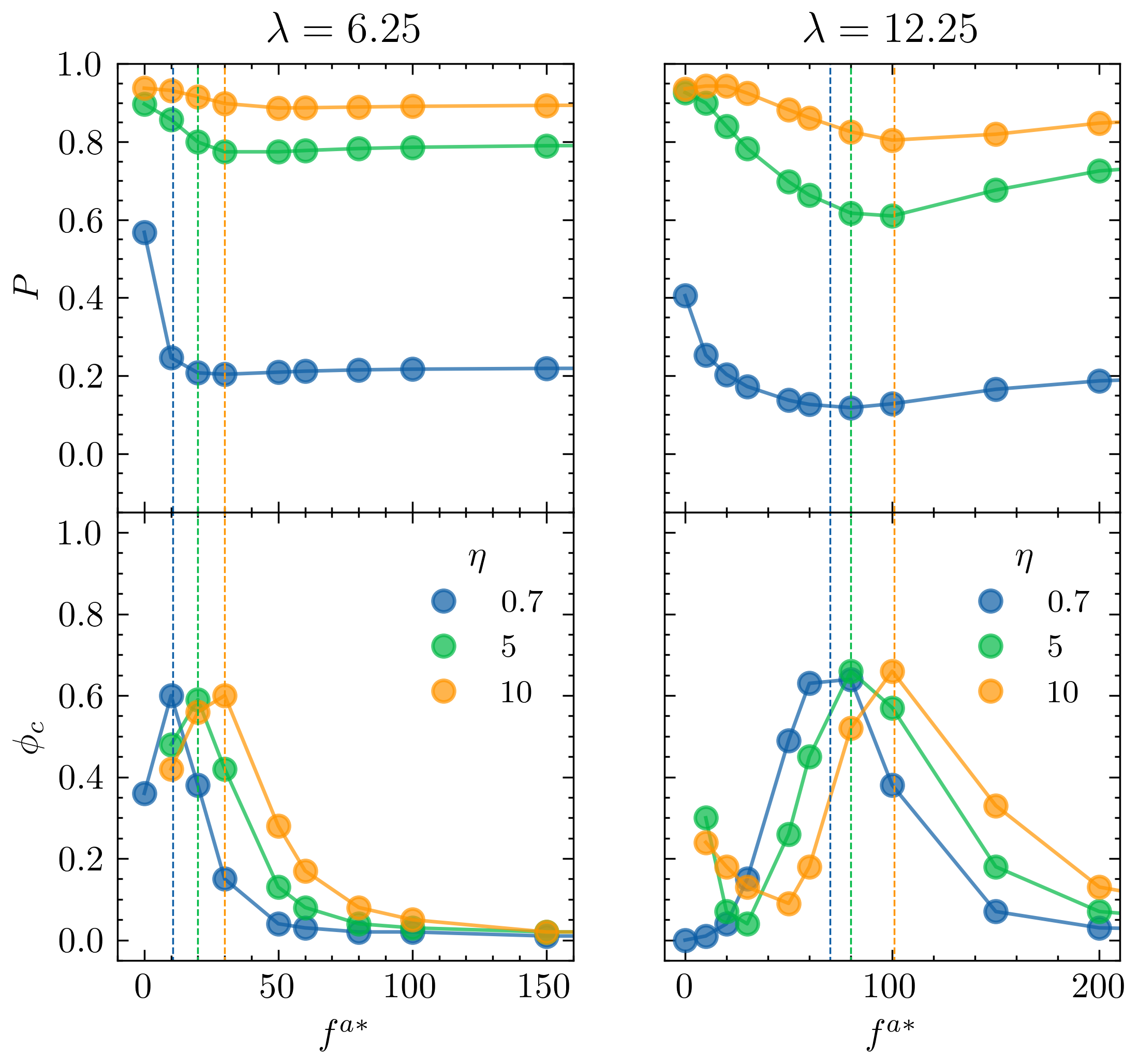}
\caption{Mean normalized polarization $P$ [Eq.~\eqref{eq:normalized_polarization}] (top panels), and mean chain fraction $\phi_c$ [Eq.~\eqref{eq:fraction_topology}] (bottom panels), as functions of the active force $f^{a*}$, at dipolar coupling strengths $\lambda=$ (\textbf{a}) 6.25 and (\textbf{b}) 12.25, and field coupling strengths $\eta = 0.7$, $5$, and $10$. Vertical dashed lines indicate the coinciding optima in both panels. The \textit{x}- and \textit{y}-axes are shared across columns and rows, respectively.}

  \begin{picture}(0,0)
  \put(-95,329){\textbf{(a)}}  
  \put(25,329){\textbf{(b)}}   
\end{picture}

       \label{fig:comp_pol_phi}
\end{figure}

We argue that the non-monotonic behavior observed in Figs.~\ref{fig:polarization_active} and \ref{fig:distributions_panels} arises \revise{from the existence of active short filamentous assemblies. As the active force increases, large interconnected branched structures break into smaller assemblies, {\it e.g.}, active chains, that frequently reorient and restructure  due to activity-induced bond breaking and reformation, causing the simultaneous depolarization of multiple dipoles. These structural correlations lead to greater polarization loss than in a gaseous state, resulting in sub-\textit{Langevin behavior}}.
In Fig.~\ref{fig:pol_loss_schematic}, we present a schematic representation of these mechanisms in two dimensions. Further increasing activity results in the onset of a \textit{gas} whose orientational responds corresponds to $P_L$, thereby giving rise to the observed non-monotonous behavior. 

This interpretation is supported by Fig.~\ref{fig:comp_pol_phi}, where we plot, in the top panels, the mean normalized polarization $P$, and in the bottom panels, the mean chain fraction $\phi_c$, as functions of $f^{a*}$ for $\lambda=$ (\textbf{a}) 6.25 and (\textbf{b}) 12.25, and for field strengths $\eta = 0.7$, $5$, and $10$. In agreement with the above suggestion, we observe that the polarization minima coincide with the chain fraction maxima, \revise{emphasizing the role of structural correlations of chain-like elements in the polarization behavior}. This maximum logically shifts to higher activities as $\eta$ increases and promotes bonding. However, although a maximum in $\phi_c$ is observed for both dipolar coupling strengths, the polarization minimum is considerably more pronounced at $\lambda = 12.25$ than at $\lambda = 6.25$.

\subsection{Orientational dynamics}
\label{subsec:orientational_dynamics}

\begin{figure}[t!]
\hspace{-1cm}
\centering
\includegraphics[width=1.05\linewidth]{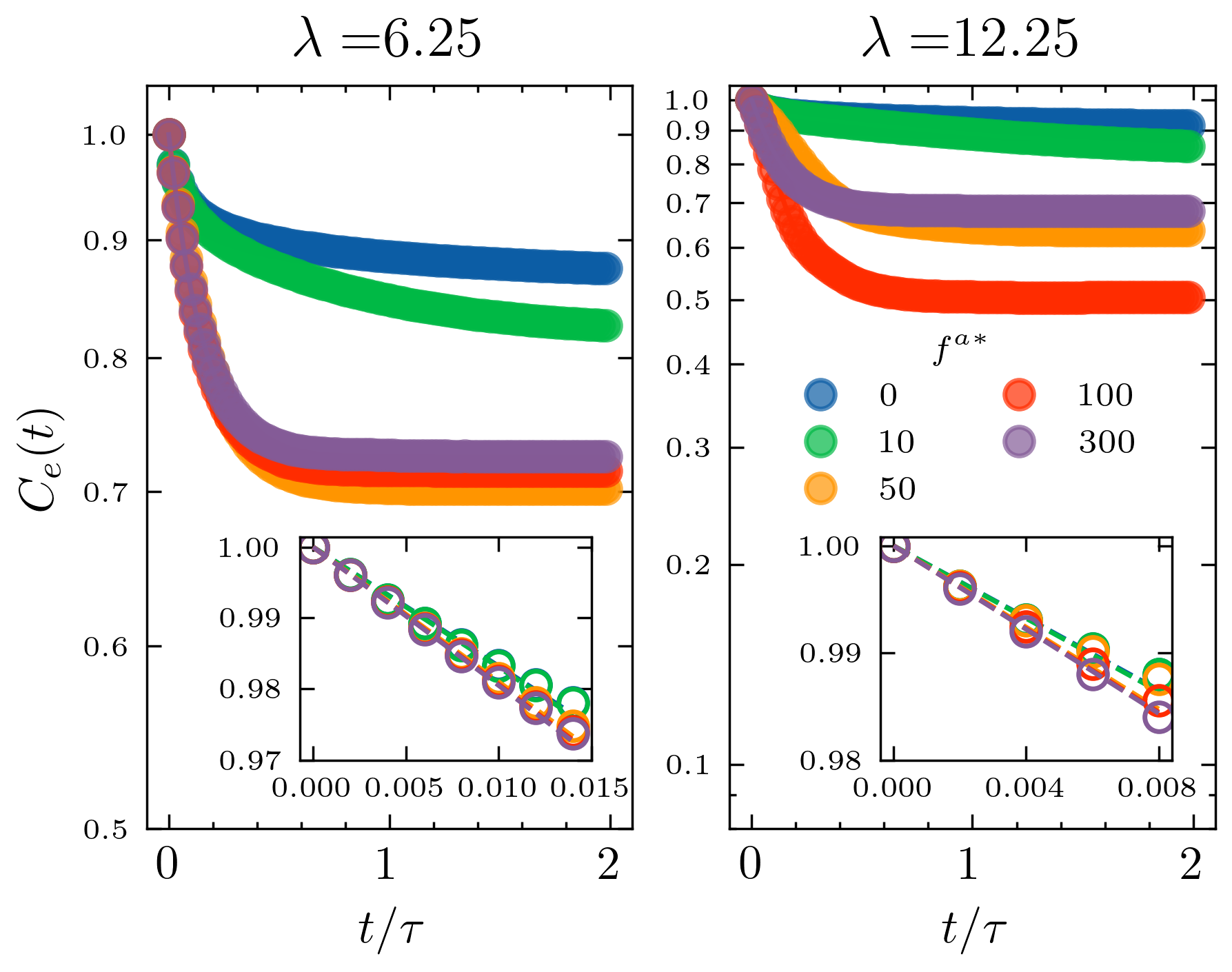}
\caption{Orientational time autocorrelation function $C_e(t)$ [Eq.~\eqref{eq:orient_tacf}] at field coupling $\eta=7$ for active forces $f^{a*}$= 0, 10, 50, 100, and 300, and dipolar coupling strengths $\lambda=$ (\textbf{a})~6.25 and (\textbf{b})~12.25. \revise{The \textit{y}-axis is in log-scale. Insets show the exponential decay of $C_e(t)$ at short-times.}}

 \begin{picture}(0,0)  
  \put(-95,279){\textbf{(a)}}
  \put(24,279){\textbf{(b)}}
\end{picture}

\label{fig:orient_tacf}
\end{figure}

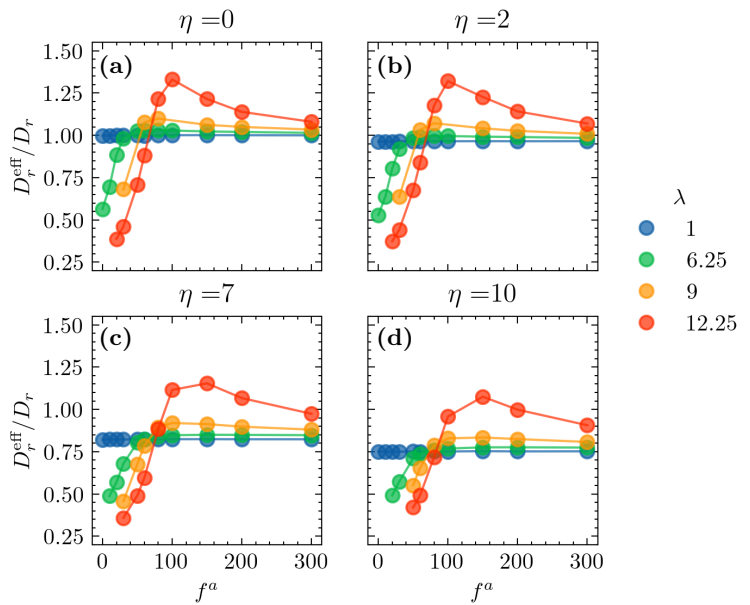
\begin{figure}[t!]
\hspace{0cm}\includegraphics[width=1.1\linewidth]{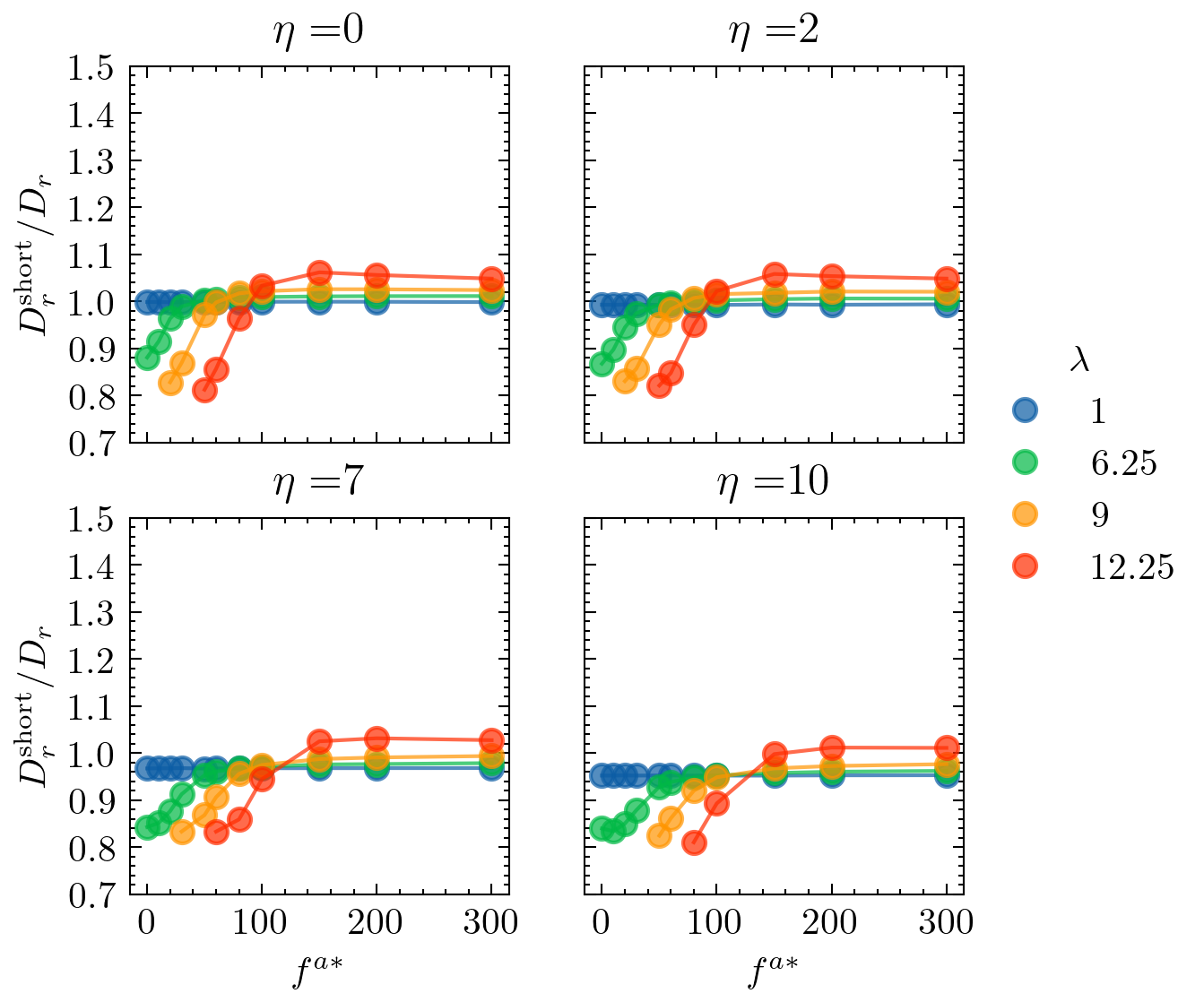}
\caption{\revise{Short-time rotational diffusion coefficient $D_r^{\text{short}}$ as a function of the active force $f^{a*}$ at field coupling strengths $\eta = \textbf{(a)}~0$, \textbf{(b)}~2, \textbf{(c)}~7, and \textbf{(d)}~10, for dipolar coupling strengths $\lambda = 1$, $6.25$, $9$, and $12.25$. The \textit{x}- and \textit{y}-axes are shared across columns and rows, respectively.}}
 \begin{picture}(0,0)  
  \put(-90,276){\textbf{(a)}}
  \put(12,276){\textbf{(b)}}
   \put(-90,175){\textbf{(c)}}
   \put(12,175){\textbf{(d)}}
\end{picture}
\label{fig:Dreff}
\end{figure}

To elucidate the non-monotonic polarization behavior highlighted in the previous section, we plot in Figs.~\refpanels{fig:orient_tacf}{a}{b} the orientational TACF $C_e(t)$ at field coupling strength $\eta=7$ and dipolar coupling strengths $\lambda=6.25$ and 12.25, for various active forces. \revise{
After an initial short-time exponential decay (see insets in Fig.~\ref{fig:orient_tacf}), the orientational TACFs exhibit a slower, non-exponential relaxation toward a plateau. By fitting $C_e(t)$ at short times  $t \lesssim 0.015$  with an exponential function of the form $\exp[-2D_r^{\text{short}}t]$, we extract the short-time rotational diffusion coefficient $D_r^{\text{short}}$. Figs.~\refpanels{fig:Dreff}{a}{d} show the short-time rotational diffusion coefficient $D_r^{\text{short}}/D_r$ as a function of the active force $f^{a*}$ for field coupling strengths $\eta = 0$, 2, 7, and 10, and various dipolar coupling strengths. For $\lambda = 1$, $D_r^{\text{short}}/D_r$ remains independent of activity across all field strengths, consistent with the system being in a gas-like state. At higher dipolar coupling strengths, where self-assembled structures emerge, $D_r^{\text{short}}/D_r$ falls below unity at low $f^{a}$ due to orientational constraints imposed by particle bonding. As the active force increases and induces bond breaking, the gradual emergence of a gaseous phase restores the rotational diffusion characteristic of free particles. 
The long-time limit of $C_e(t)$, related to the mean polarization via $\lim_{t \to \infty} C_e(t) = |\langle \hat{\mathbf{e}}_i \rangle|^2$, varies non-monotonically with the active force $f^{a*}$. It first decreases to a minimum before approaching a saturation value, with the minimum more pronounced for $\lambda = 12.25$ than for $\lambda = 6.25$. This trend corroborates the behavior of the polarization discussed in Sec.~\ref{subsec:polarization}, suggesting that these variations reflect changes in the structural state of the system.}

\subsection{Effect of external field on polymerization and network connectivity
}
\label{subsec:polymerization_and_percolation}

\begin{figure}[h!] 
     \includegraphics[width=1.02\linewidth]{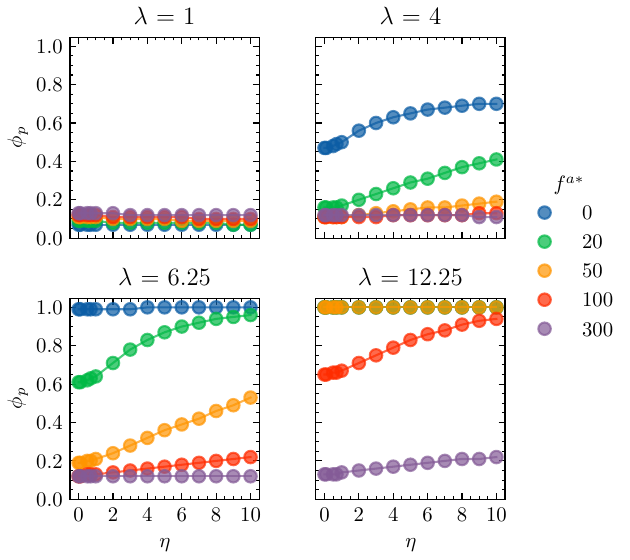}

     \caption{Polymerization degree $\phi_p$ [Eq.~\eqref{eq:phi_pol}], as a function of the field coupling strength $\eta$ at dipolar coupling strengths $\lambda=$ (\textbf{a}) 1, (\textbf{b}) 4, (\textbf{c}) 6.25, and (\textbf{d}) 12.25, and for active forces $f^{a*}$= 0, 20, 50, 100, and 300. The \textit{x}- and \textit{y}-axes are shared across columns and rows, respectively.}
  \begin{picture}(0,0)
  \put(-91,274){\textbf{(a)}}  
  \put(7,274){\textbf{(b)}}   
   \put(-91,163){\textbf{(c)}}   
   \put(7,163){\textbf{(d)}}   
\end{picture}
    \label{fig:polymerization}
\end{figure}
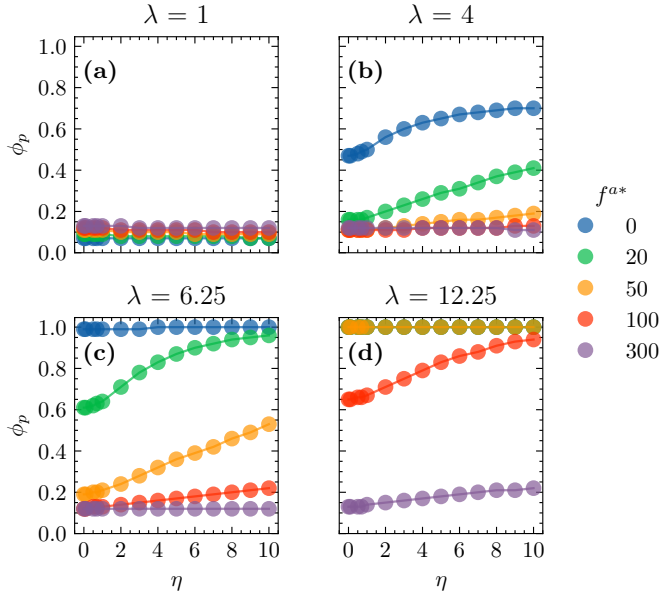

Given the highlighted interplay between structure and response to the external field, we examine in this section how the field influences structural changes. To this end, we plot in Figs.~\refpanels{fig:polymerization}{a}{d} the polymerization degree $\phi_p$ as a function of $\eta$ for dipolar coupling strengths $\lambda=$ 1, 4, 6.25, and 12.25, and various active forces. At the lowest value of $\lambda = 1$, where the systems remain in a \textit{gas} state, polymerization is largely independent of the field strength, showing only a slight increase with $f^{a*}$ due to enhanced two-body collisions \cite{Tailleur, Redner}. For higher values of the dipolar coupling strength $\lambda \geq 4$, polymerization increases significantly with $\eta$, provided that the systems are not fully polymerized in the zero-field limit and that the active force is sufficiently low compared to the strength of the dipolar forces and field strength. As previously mentioned in Sec.~\ref{sec:passive_limit}, this mechanism arises from the alignment field stabilizing head-to-tail bonding configurations, in competition with activity, just as observed in the zero-field limit~\cite{Liao, Kelidou}.

\begin{figure}[b!]
 \includegraphics[width=9cm]{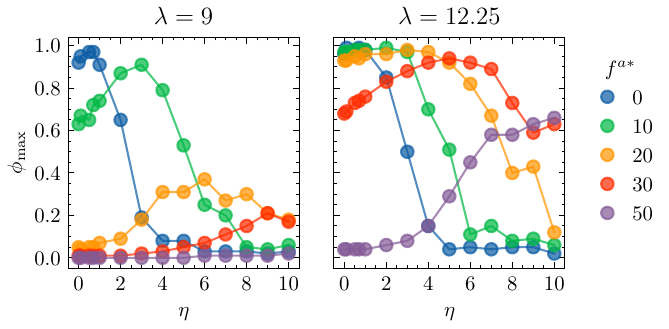}
    \caption{Mean fraction of particles in the largest cluster $\phi_{\text{max}}$ [Eq.~\eqref{eq:phi_max}], as a function of the field coupling strength $\eta$ at $\lambda=$ (\textbf{a}) 9, and (\textbf{b}) 12.25, and for active forces $f^{a*}$= 0, 10, 20, 30, and 50. The \textit{y}-axis is shared by both subfigures.}
\begin{picture}(0,0)
  \put(-22,169){\textbf{(a)}}  
  \put(82,169){\textbf{(b)}}    
\end{picture}
  		\label{fig:perc_deg_field}
\end{figure}
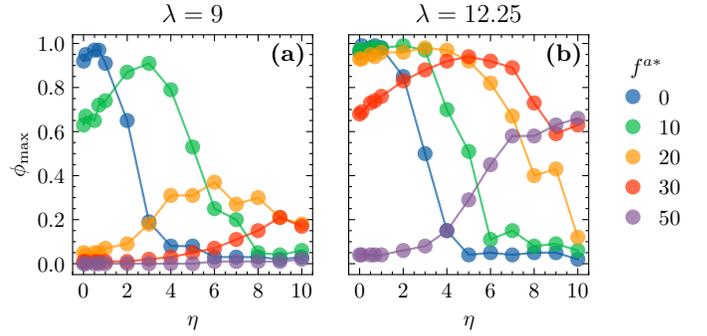

Next, we investigate the effect of an external field on the connectivity properties by plotting the mean fraction of particles in the largest cluster $\phi_{\text{max}}$ as a function of $\eta$ for $\lambda = 9$ and $12.25$ and various active forces, as shown in Figs.~\refpanels{fig:perc_deg_field}{a}{b}. In the passive limit, \textit{percolated networks} are rapidly destroyed as $\eta$ increases, giving way to a polarized fluid of aligned strings [Fig.~\ref{fig:passive_snapshots}], as indicated by the sharp drop in $\phi_{\text{max}}$. For active systems, the interplay between activity and external field leads to a more complex, non-monotonic dependence of $\phi_{\text{max}}$ on $\eta$. We observe that weak to moderate fields enhance network connectivity, with $\phi_{\text{max}}$ either stabilizing or increasing to a maximum before declining at higher $\eta$. This effect is most pronounced at $\lambda = 9$ and $f^{a*}$= 10, where the fraction of particles in the largest cluster increases from $\phi_{\text{max}} < 0.7$ in the zero-field limit, indicating a system of many independent active structures, to $\phi_{\text{max}} > 0.7$, signaling the formation of a \textit{percolated network}. At stronger fields, connectivity is eventually lost, giving rise to a \textit{polarized string fluid} with $\phi_{\text{max}} \rightarrow 0$. 

\begin{figure}[t!]
\includegraphics[width=1.05\linewidth]{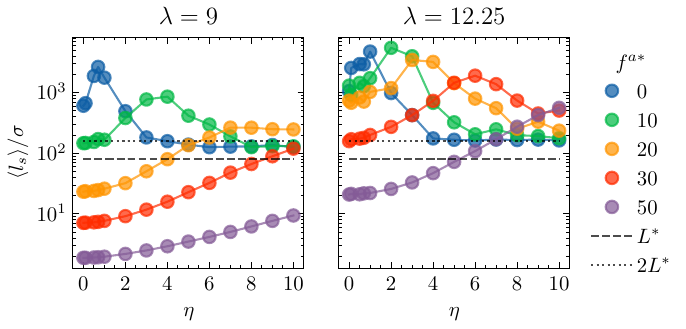}
 \caption{Average cluster size $\langle l_s \rangle$ [Eq.~\eqref{eq:ave_size}], as a function of the field coupling strength $\eta$ at dipolar coupling strengths $\lambda=$ (\textbf{a}) 9, and (\textbf{b}) 12.25, for active forces $f^{a*}$= 0, 10, 20, 30, and 50. The symbol $L^{*}$ corresponds to the simulation box's edge length. The \textit{y}-axis is shared by both subfigures.}
 \begin{picture}(0,0)
  \put(-21,175){\textbf{(a)}}  
  \put(78,175){\textbf{(b)}}    
\end{picture}
   		\label{fig:ave_clust_size}
\end{figure}
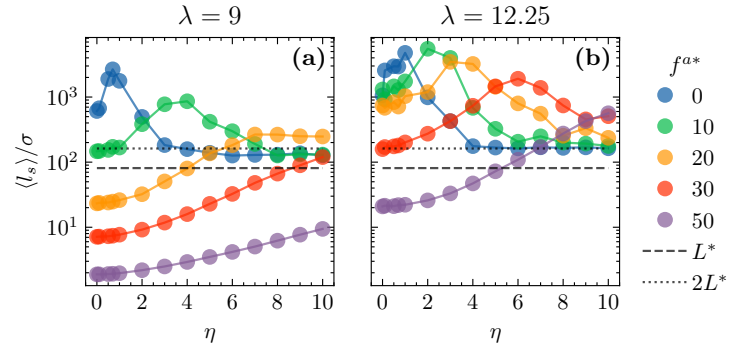

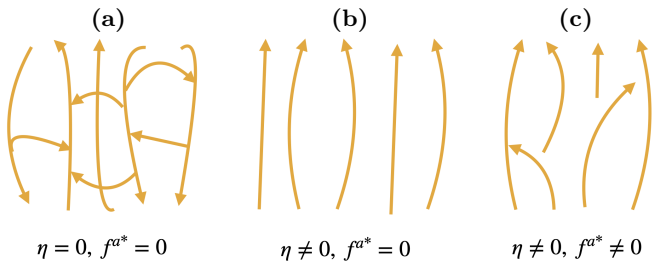
\begin{figure}[b!]
\includegraphics[width=1.05\linewidth]{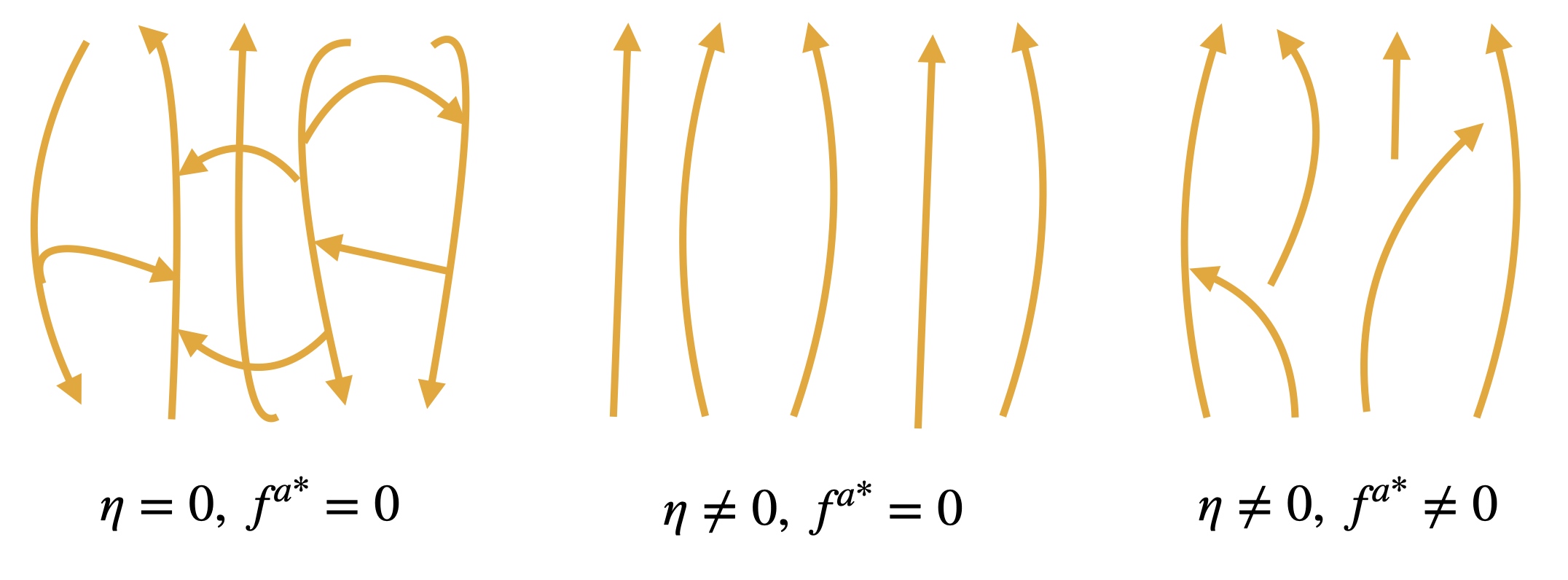}
   \caption{Schematics where arrows represent chains of bonded dipolar particles: \textbf{(a)} passive \textit{percolated network} in the zero field limit ($\eta = 0$), with random orientations of the network's branches; \textbf{(b)} isolated passive polarized columnar clusters under an external field ($\eta > 0$); \textbf{(c)} active polarized columnar clusters under an external field ($\eta > 0$), which connect as a result of bond-breaking.}
 \begin{picture}(0,0)
  \put(-85,192){\textbf{(a)}}  
  \put(6,192){\textbf{(b)}} 
\put(92,192){\textbf{(c)}} 
\end{picture}
\label{fig:scheme_percolation}
\end{figure}

We plot in Figs.~\refpanels{fig:ave_clust_size}{a}{b} the average cluster size, $\langle l_s \rangle$, as a function of $\eta$ for $\lambda = 9$ and $12.25$ and various active forces. In the passive limit, the average cluster size at both $\lambda$ values initially increases with  $\eta$ up to a maximum, before converging toward values between $L^{*}$ and $2L^{*}$, signaling the formation of columnar pair-bundles of chains \cite{Weis3, Weis4}. We note that the initial slight increase in $\langle l_s \rangle$ at low field for $f^{a*}$= 0 is somewhat artificial, as it results from the breakage and inclusion of small rings, coexisting with the \textit{percolated network} at zero-field, into the main structure. Similar trends to those in Fig.~\ref{fig:perc_deg_field} are observed at low to moderate activity levels, where weak fields induce a substantial increase in $\langle l_s \rangle$, indicating that isolated active structures connect to form larger aggregates. We suggest that this is made possible by the coupling of enhanced activity-induced bond-breaking and polarization. At low $\eta$ values, the constant reorganization of active structures is only loosely constrained in the field direction, favoring encounters between independent aggregates and thereby facilitating connectivity. At high $\eta$ values, the alignment torque exerted by the field strongly orients the dipoles along the field direction, suppressing reorganization and leading to the observed leveling off in $\langle l_s \rangle$. A schematic representation of this mechanism is presented in Fig.~\ref{fig:scheme_percolation},  attributing the greater resilience of active networks to the cooperative effects of activity and external field.

\subsection{Bond dynamics}
\label{subsec:bond_dynamics}

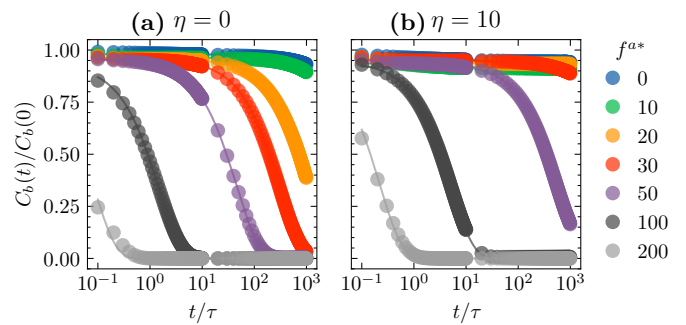
\begin{figure}[h!]
\centering
\includegraphics[width=1.12\linewidth]{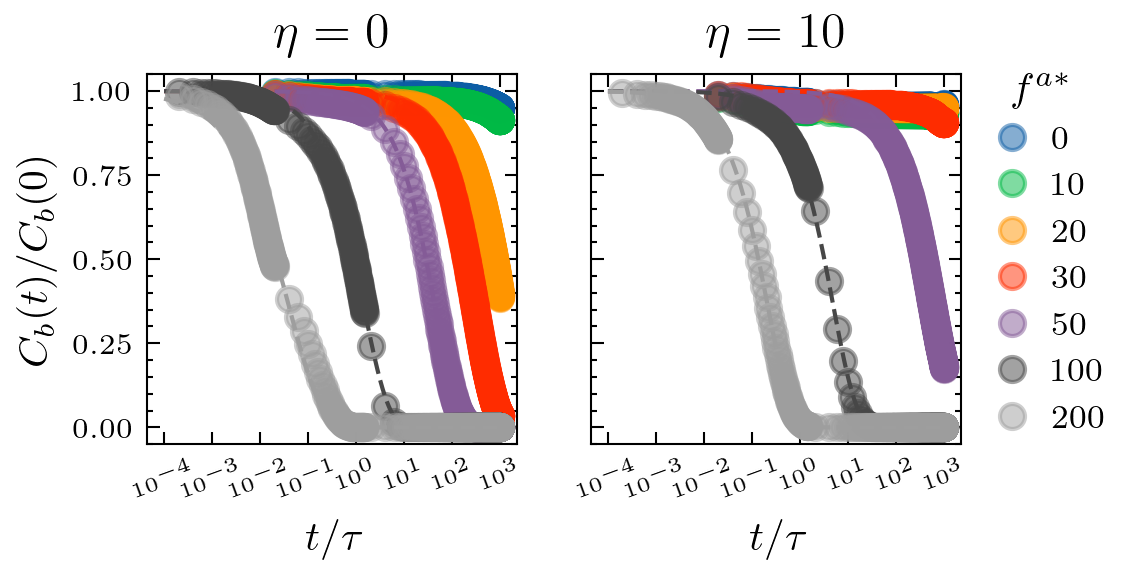}
\caption{\revise{Normalized bond TACF $C_b(t)/C_b(0)$ [Eq.~\ref{eq:bond_tacf}], at $\lambda = 12.25$ for field-strength $\eta=$ (\textbf{a}) 0, and (\textbf{b}) 10, and activities $f^{a*}$= 0, 10, 20, 30, 50, 100, and 200. Dashed lines connecting the markers represent stretched-exponential fits. The \textit{y}-axis is shared by both subfigures.}}
 \begin{picture}(0,0)
  \put(-70,203){\textbf{(a)}}  
  \put(34,203){\textbf{(b)}}    
\end{picture}
\label{fig:bond_TACF}
\end{figure}

In this last section, in light of the highlighted phenomena, we quantify the interplay between activity and external field on the bonds formed by dipolar particles. To this end, we plot in Figs.~\refpanels{fig:bond_TACF}{a}{b} the normalized bond TACF for $\eta = 0$ and $10$ at $\lambda = 12.25$ for various activity levels. In the zero-field limit, consistent with our previous findings in \cite{Kelidou}, we find that as the active force increases, the bond TACFs progressively decay faster, reflecting the onset of more transient structures. For active forces $f^{a*} \leq 20$, where the system maintains a \textit{percolated network} structure, the decay of $C_b(t)$ is observed to be faster compared to the passive case, indicating a more dynamic and time-reconfiguring network structure. At $\eta=10$,  a similar trend is observed with activity, although activities up to $f^{a*}$= 30 are characterized by a rather non-decaying $C_b(t)$. This is a result of the external field which by favoring "head-to-tail" alignments delays the decay of $C_b(t)$ to larger times. 

\begin{figure}[t!]
\includegraphics[width=0.8\linewidth]{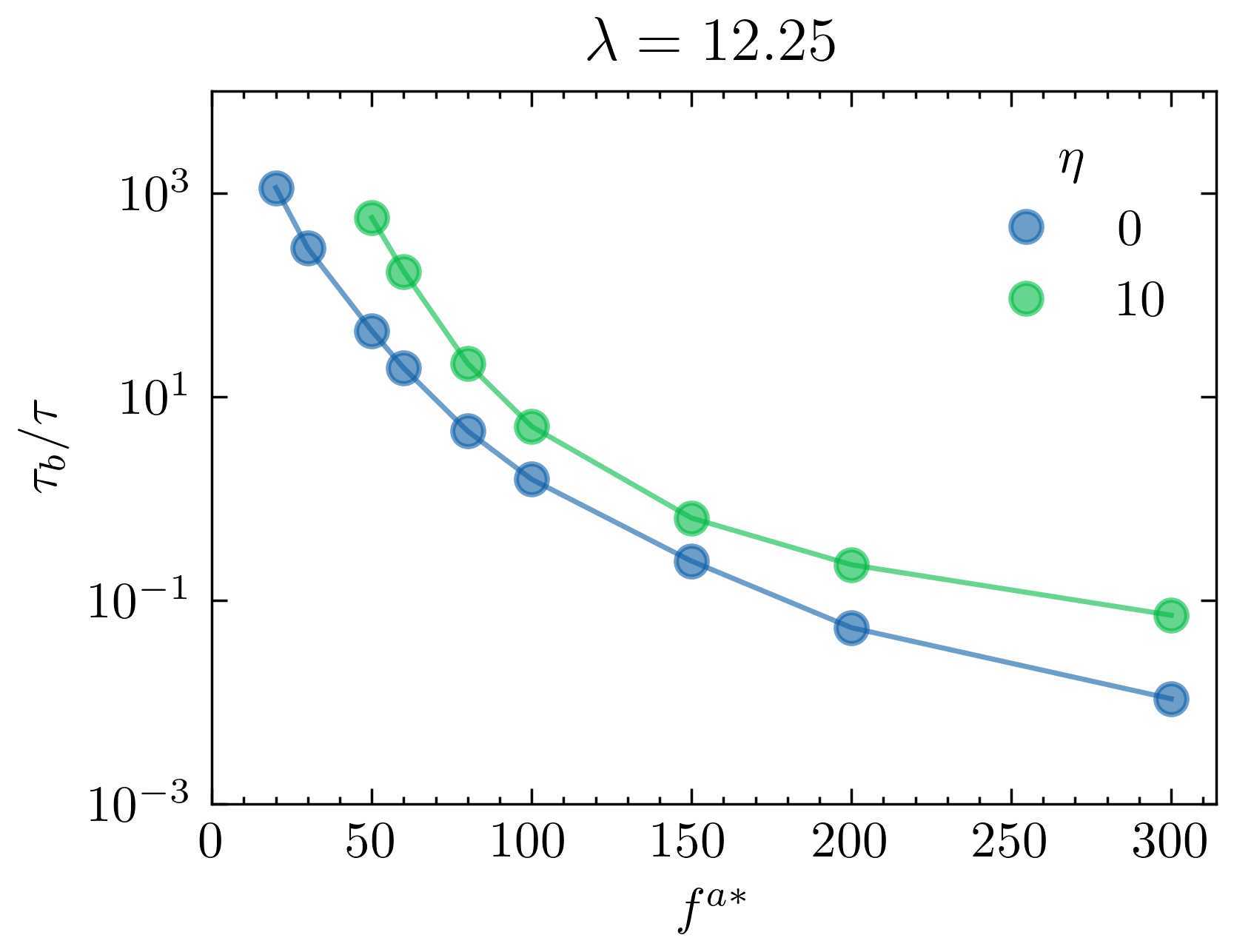}
\caption{\revise{Mean bond lifetime  $\tau_b$ as a function of the active force $f^{a*}$ at $\lambda=12.25$, for field coupling strengths $\eta=$ 0, and 10.}}

\label{fig:bond_timelife}
\end{figure}

\revise{We find that the $C_b(t)$ curves can be well described by a stretched-exponential decay of the form $\exp[-(t/\tau_K)^{\beta}]$, where $\tau_b = \tau_K\Gamma(\beta^{-1})/\beta$ can be interpreted as the mean bond lifetime \cite{Elton}, with $\Gamma$ the well-known gamma-function}. The evolution of $\tau_b$ with activity at $\lambda = 12.25$ is presented in Fig.~\ref{fig:bond_timelife} for field coupling strengths $\eta = 0$ and $10$. Consistent with previous observations, increasing $f^{a*}$ leads to a substantial decrease in the mean bond lifetime. \revise{Conversely, $\tau_b$ values are systematically larger at $\eta = 10$ than in the zero-field case, which corroborates the arguments made in Sec.~\ref{subsec:polymerization_and_percolation} that the interplay between activity and the external field consists of a competition between two antagonistic effects for bond stability, with activity weakening bonds and the external field favoring their formation.}

\section{Conclusions}
\label{sec:conclusions}

In this work, we investigated the response of low-density three-dimensional assemblies of active dipolar particles to an external field using Brownian dynamics simulations, systematically varying their dipolar coupling strength and activity level. In our previous work in the zero-field limit \cite{Kelidou}, we highlighted 
emergence of two new non-equilibrium structures at low densities. They include an isotropic fluid of active strings and an active \textit{percolated network} that undergoes frequent dynamical reconfigurations, emerging at low to moderate activity and sufficiently strong dipolar coupling.

Under an external field, we find that active assemblies exhibit a much richer phenomenology than their passive counterparts. In passive systems, the zero-field structures simply polarize along the direction of the field, leading to (i) an increased polymerization at low dipolar coupling, and (ii) the formation of independent columnar clusters at higher dipolar coupling, signifying the breakdown of the network structures. In contrast, in active systems, the orientational constraint imposed by the external field and self-propulsion can act cooperatively to favor connectivity. In particular, we find that (i) low activity levels extend the range of fields over which a \textit{percolated network} persists, and (ii) at strong dipolar coupling strengths, a weak external field can induce system-spanning networks at intermediate activity levels. These results demonstrate the greater robustness of active networks to weak external fields compared to their passive counterparts, a consequence of continuous activity-induced dynamical rearrangements, where moderate orientational constraints enhance the likelihood of structural encounters during reorganization. Conversely, high fields disrupt network connectivity, leading to the formation of \textit{polarized string fluids}. In addition, contrary to the passive case, active systems with strong dipolar coupling strengths exhibit a lower polarization than a system  of independent dipoles, \revise{which we attribute to the presence of active self-assembled structures. In the large-activity regime, where binding is suppressed, the system forms a dipolar active gas with a polarization response following the \emph{Langevin} function.}  

\revise{In the ongoing pursuit of strategies to control and program the self-assembly of active particles, our findings highlight the effectiveness of external fields in directing the structural and orientational organization of self-propelled dipolar systems. These results provide valuable guidelines for manipulating the collective behavior of magnetotactic bacteria~\cite{magnetotactic_bacteria_Bazylinski,MTB_24} via external fields. Moreover, recent experimental realizations of active dipolar particles in both two- and three-dimensions~\cite{Yan,Chao,magnetic_roller,Reyes2023magnetic}, where self-propulsion arises from induced charge electrophoresis~\cite{Squires}, simultaneously generating electric dipoles, exhibit closely related dynamics. 
We anticipate similar phenomenology for active dipolar colloids driven by electrophoresis. In such systems, however, the presence of ions screens the electrostatic potential, making the dipolar interactions effectively short-ranged. Moreover, one must consider the intrinsic coupling between dipole–dipole interactions and particle activity induced by the external field. Lastly, the interplay between hydrodynamic and dipolar interactions is expected to play an important role for dipolar microswimmers at appropriate concentration and dipolar strength regimes~\cite{Guzman2016fission,koessel1,Koessel2}.}

\revise{As a promising avenue for future research, exploring the hysteresis behavior of active \textit{percolated networks} could open new possibilities for designing the next generation of smart, reconfigurable materials. In particular, the combination of activity and external fields may serve as an effective mechanism for memory reset, enabling controllable self-assembly and disassembly cycles. Another open question concerns the relationship  between connectivity and rigidity percolation in active networks, and how it governs their emergent mechanical properties. More broadly, extending the present study to higher densities, where passive dipolar particles are known to spontaneously develop orientational order~\cite{Stevens,Weis2}, could reveal collective flocking states analogous to those observed in two-dimensional active dipolar systems~\cite{Liao}.}

\begin{acknowledgments}
This work used the Dutch national e-infrastructure with the support of the SURF Cooperative using Grant No. EINF-8801. 
\end{acknowledgments}

\appendix

\section{Analysis details}
\label{ap:analysis_details}

\subsection{Clustering analysis}
\label{apsub:clustering_analysis}

\revise{Dipolar particles in 3D can self-assemble into chains, rings, and \textit{percolated network} structures in the low-density regime studied here \cite{Stevens, Weis, Weis2, Sciortino,Kelidou}. To classify the states observed in simulations, we perform a clustering analysis of the configurations based on a simple distance criterion~\cite{Kelidou,Liao}: two particles are considered bound if their center-to-center distance $r_{ij}$ is smaller than a threshold cutoff, $r_b$. The distance criterion was set at $r_b^*$= 1.2, based on the first minimum of the radial distribution function,  almost insensitive to variations in $\lambda$, $f^{a*}$, or $\eta$. We used \textit{k-d} trees \cite{SciPy} for an efficient nearest neighbor search.} 
Identified clusters are then sorted according to three topologies: \textit{chains} are clusters with more than two particles, in which there exist exactly two particles with one neighbor and all the other particles have two neighbors; \textit{rings} are clusters in which every particle has exactly two neighbors; \textit{branched structures} are clusters in which at least one particle has more than two neighbors. 

We define the probability $P_{\alpha}(n)$ of finding a cluster of size $n$ with topology $\alpha$, with $\alpha = c$ for \textit{chains}, $\alpha = r$ for \textit{rings}, $\alpha = b$ for \textit{branched structures}, and $\alpha = s$ refers to any self-assembled topology, as
\\
\begin{equation}
\label{eq:3.10}
P_{\alpha}(n) =\left \langle \frac{N_{\alpha}(n)}{\sum_{n\geq n_{\alpha}^{\text{min}}}N_{\alpha}(n)}\right \rangle, 
\end{equation}
\\
\noindent where $N_{\alpha}(n)$ is the number of clusters of type $\alpha$ with length $n$ particles, $\langle...\rangle$ refers to an average over relevant configurations, and the minimum length $n_{\alpha}^{\text{min}}$ equals 3 for topology $c$ and $r$, 4 for topology $b$, and 1 for topology $s$. From equation (\ref{eq:3.10}) we can calculate the average size of clusters with topology $\alpha$ as 
\begin{equation}
\langle l_{\alpha} \rangle = \sum_{n \geq n_{\alpha}^{\text{min}}} n P_{\alpha}(n). 
\label{eq:ave_size}
\end{equation}
Further, we define the mean fraction of particles belonging to structures with topology $\alpha$ as 
\begin{equation}
\label{eq:fraction_topology}
\phi_{\alpha} = \left \langle\frac{\sum_{n\geq n_{\alpha}^{min}} nN_{\alpha}(n)}{N}\right \rangle.
\end{equation}
\\
 The polymerization degree, \textit{i.e.} the fraction of particles belonging to any cluster with size $n\geq 2$ regardless of its topology, is given by
\begin{equation}
\label{eq:phi_pol}
\phi_p = 1 - \phi_1
\end{equation}
where $\phi_1 = \langle N_s(n=1)/N\rangle$ is the fraction of monomers in the system. Lastly, in order to assess the existence of a \textit{percolated network} in the system, we determine the mean number of particles in the largest cluster, whatever its topology,
\begin{equation}
\label{eq:phi_max}
\phi_{\text{max}} =  \left \langle\frac{\max\limits_{s} \{n\}
}{N}\right \rangle
\end{equation}
where $\max\limits_{s} \{n\}$ denotes the largest size $n$ among all clusters regardless of their specific topology.

\subsection{Angular distribution and mean polarization}
\label{apsub:polarization}
Under the influence of an external field, the particles' dipoles are likely to align with the field direction, resulting in polarized states. To quantify this effect, we compute (i) the probability distribution $\psi$ of finding a dipole forming an angle $\theta$ with the field axis, where $\theta = \arccos{(\hat{\mathbf{e}}_i \cdot \hat{\mathbf{B}})}$, and (ii) the mean normalized polarization, defined as
\begin{equation}
P = \Big\langle \frac{1}{N} | \sum_{i=1}^N \hat{\mathbf{e}}_i | \Big\rangle,
\label{eq:normalized_polarization}
\end{equation} 
In the case of an assembly of non-interacting dipoles with a continuous distribution of dipole energy states, the angular probability distribution $\psi_L$ is obtained from a \textit{Boltzmann} distribution $\sim \exp{(-\beta U_{\text{df}})}$, and is given by \cite{Langevin}
\begin{equation}
    \psi_L= \frac{\eta}{2 \sinh{\eta}} \exp{(\eta \cos \theta}),
\label{eq:langevin_distribution}
\end{equation}
The corresponding mean normalized polarization is the well-known \textit{Langevin} function, whose functional form in three dimensions is \cite{Langevin}
\begin{equation}
P_L= \coth(\eta) - \frac{1}{\eta}
\label{eq:langevin_polarization}
\end{equation}

\subsection{Dynamical features}
\label{apsub:dynamical_features}

To quantify the bond dynamics, we compute a bond time autocorrelation function (TACF) which measures the probability that a bond existing at time $t'$ still exists at time $t'+t>t'$, quantifying how fast the structures rearrange. This is done by assigning a bond variable \(n_{ij}(t)\) at each time \(t\) for every pair of particles \(i\) and \(j\), as described in \cite{DelGado}. This variable is set to 1 if particles \(i\) and \(j\) are neighbors according to the distance criterion introduced in Appendix~\ref{apsub:clustering_analysis}, and 0 otherwise. The bond TACF is then calculated as
\begin{equation}
\label{eq:bond_tacf}
    C_b(t)=\frac{2}{N(N-1)} \sum_{i \neq j}\langle n_{ij}(t'+t) \cdot n_{ij}(t') \rangle_{t^{'}}
\end{equation} 
where $\langle...\rangle_{t'}$ indicates averaging over all possible starting times $t'$.

To characterize the orientational dynamics of dipolar active particles, we compute the time autocorrelation function of the unit orientation vector $\hat{\mathbf{e}}_i$, 
\begin{equation}
    C_e(t) = \frac{1}{N} \sum_{i=1}^N \left\langle \hat{\mathbf{e}}_i(t'+t) \cdot \hat{\mathbf{e}}_i(t') \right\rangle_{t'}
\label{eq:orient_tacf}
\end{equation}
For non-interacting Brownian particles, $C_e(t)$ decays as $\exp(-2D_rt)$, where the rotational diffusion coefficient $D_r$ is defined in Eq.~\eqref{eq:3.6}.

\section{Spatial ordering}
\label{ap:spatial_ordering}

\begin{figure}[h!]
    \centering
    \hspace{-0.5cm}
    \includegraphics[width=1\linewidth]{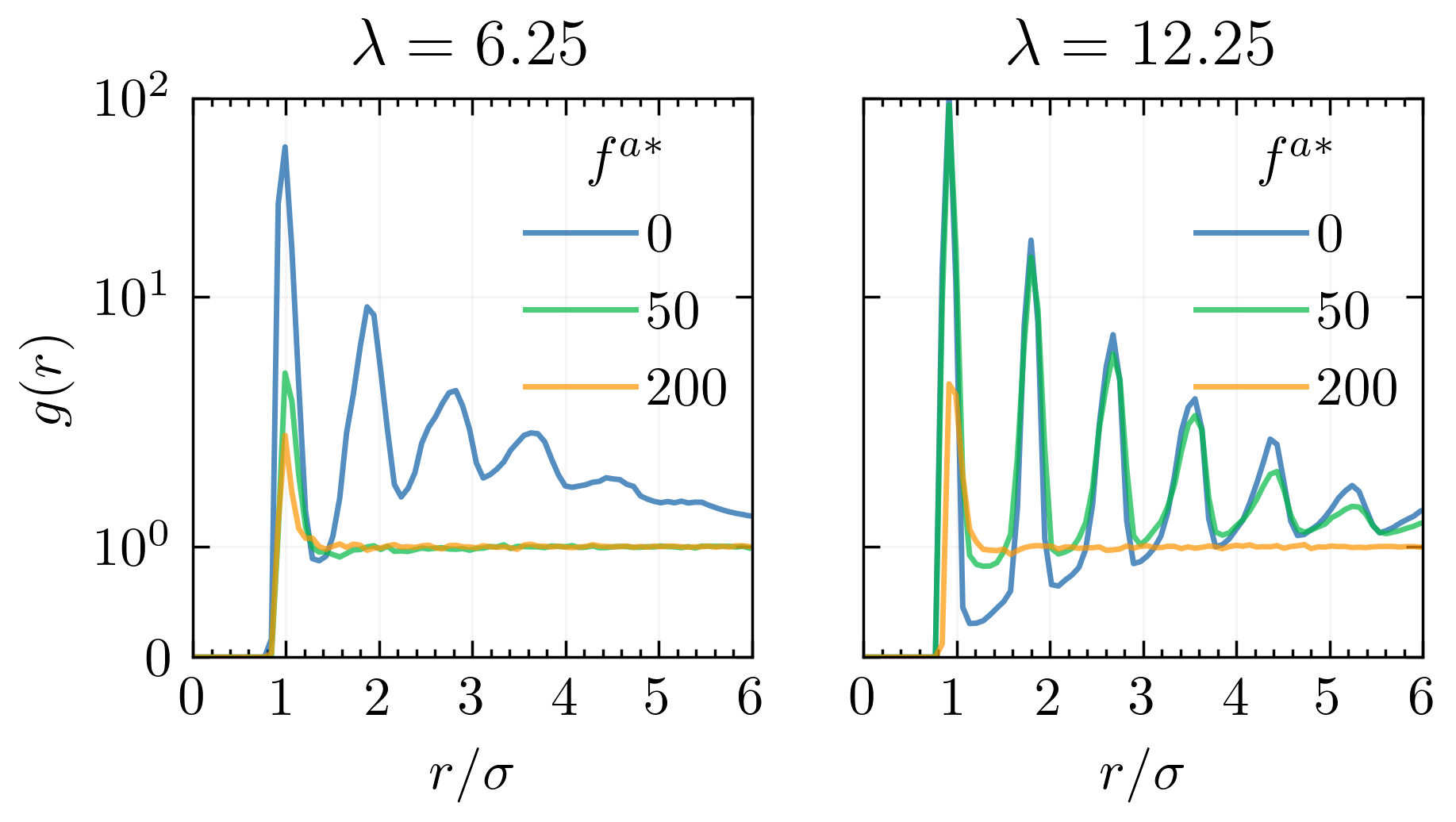}
    \caption{Isotropic radial distribution function $g(r)$ [Eq.~\eqref{eq:iso_rdf}], in the zero-field limit $\eta = 0$ for dipolar coupling strengths $\lambda=$ \textbf{(a)} 6.25 and \textbf{(b)} 12.25, and active forces $f^{a*}$= 0, 50, and 200.}

     \begin{picture}(0,0)
  \put(-84,183){\textbf{(a)}}  
  \put(27,183){\textbf{(b)}}    
\end{picture}
    
    \label{fig:rdf_iso}
\end{figure}

\revise{On Fig.~\refpanels{fig:rdf_iso}{a}{b}, we show the isotropic radial distribution function}
\begin{equation}
g(r) = \frac{1}{N\rho}\frac{\langle \sum_{i\neq j} \delta(r_i - r_j)\rangle}{4\pi r^2},
\label{eq:iso_rdf}
\end{equation}
\revise{computed in the zero-field limit ($\eta=0$) for dipolar coupling strengths $\lambda = 6.25$ and $12.25$, and activities $f^{a*} = 0$, $50$, and $200$. As $f^{a*}$ increases, $g(r)$ evolves from displaying many regularly spaced peaks, characteristic of chain-like self-assembled structures, to a single peak at $r/\sigma = 1$, signaling the onset of a \textit{gas} phase. It is noteworthy that the first-neighbor peak does not shift substantially below $r/\sigma = 1$ with increasing $f^{a*}$, confirming that the interactions remain quasi-hardcore.}

We briefly discuss the shape of two direction-dependent radial distribution functions (RDFs) measuring the positional correlations parallel ($g_{\parallel}$) and perpendicular ($g_{\perp}$) to the external field direction $\hat{\mathbf{B}}$. They are defined as \cite{Mandle}
\begin{align}
g_{\parallel}(r_{\parallel}) &= \frac{1}{N\rho} \frac{\left\langle \sum_{i \neq j} \delta\left(r_{\parallel}-  \lVert \mathbf{r}_{ij}\cdot \hat{\mathbf{B}} \rVert\right)\right\rangle}{2\pi (\sigma_{\parallel}/2)^2} \label{eq:para_rdf} \\[0.5em] 
g_{\perp}(r_{\perp}) &= \frac{1}{N\rho} \frac{\left\langle \sum_{i \neq j} \delta\left(r_{\perp}- \lVert \mathbf{r}_{ij} \times \hat{\mathbf{B}} \rVert\right)\right\rangle}{2\pi r_{\perp} \sigma_{\perp}} \label{eq:perp_rdf}
\end{align}
where $i$ and $j$ refer to particle indices, $r_{\parallel}$ and $r_{\perp}$ are, respectively, the distances from particles’ centers parallel and perpendicular to the field, and $\sigma_{\parallel}$ and $\sigma_{\perp}$ are, respectively, the diameter and height of cylindrical sampling volumes oriented parallel to $\hat{\mathbf{B}}$.

\begin{figure}[b]
\centering
\includegraphics[width=1\linewidth]{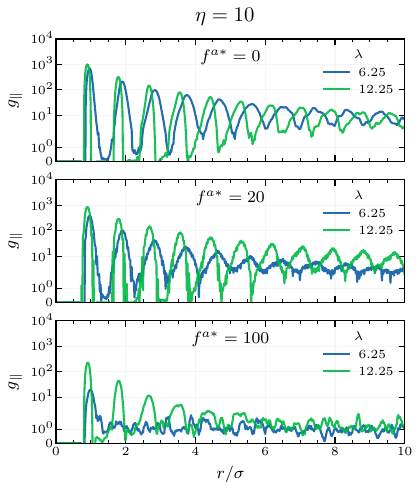}
\caption{
Parallel radial distribution function $g_{\parallel}$ [Eq.~\eqref{eq:para_rdf}], 
at field coupling strength $\eta = 10$ for dipolar coupling strengths 
$\lambda = 6.25$ and 12.25, and active forces $f^{a*}$ = 
(\textbf{a}) 0, (\textbf{b}) 20, and (\textbf{c}) 100. 
A value of $\sigma_{\parallel} = 0.5\sigma$ was used, 
and data were smoothed using a Savitzky–Golay filter \cite{Savitzky}. 
The \textit{x}-axis is shared by all subfigures.
}
\label{fig:rdf_para}

\begin{picture}(0,0)
  \put(-87,340){\textbf{(a)}}  
  \put(-87,257){\textbf{(b)}}   
  \put(-87,176){\textbf{(c)}}   
\end{picture}
\end{figure}

\begin{figure}[t!]
\centering
\includegraphics[width=1\linewidth]{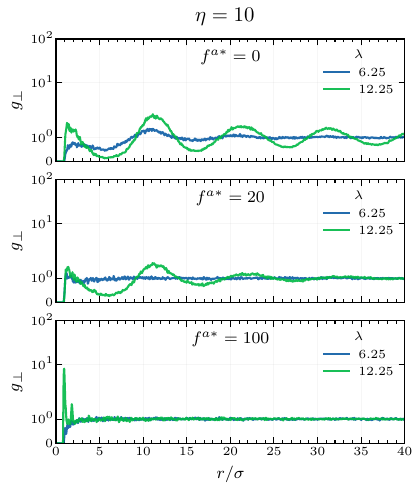}
\caption{
Perpendicular radial distribution function $g_{\perp}$ [Eq.~\eqref{eq:perp_rdf}], 
at field coupling strength $\eta = 10$ for dipolar coupling strengths 
$\lambda = 6.25$ and 12.25, and active forces $f^{a*}$ = 
(\textbf{a}) 0, (\textbf{b}) 20, and (\textbf{c}) 100. 
A value of $\sigma_{\perp} = 0.5\sigma$ was used, 
and data were smoothed using a Savitzky–Golay filter \cite{Savitzky}. 
The \textit{x}-axis is shared by all subfigures.
}
\label{fig:rdf_perp}

\begin{picture}(0,0)
  \put(-87,340){\textbf{(a)}}  
  \put(-87,257){\textbf{(b)}}   
  \put(-87,176){\textbf{(c)}}   
\end{picture}
\end{figure}

Figs.~\refpanels{fig:rdf_para}{a}{c} shows the parallel RDF $g_{\parallel}$ at the highest investigated field strength, $\eta = 10$, for activities $f^{a*}$= 0, 20, and 100, and dipolar coupling strengths $\lambda = 6.25$ and 12.25. In the passive limit, the  strings, spanning  the entire box length, highlighted in the main text, are well-distinguishable by several regularly spaced peaks in $g_{\parallel}$, signaling linear segments of particles oriented along the field. Upon increasing activity, columnar clusters shorten as long-distance peaks in $g_{\parallel}$ are progressively lost. This happens more quickly at $\lambda=6.25$ than $\lambda=12.25$, consistent with the trends discussed in Sec. \ref{subsec:polymerization_and_percolation}.

The perpendicular RDF $g_{\perp}$ is shown in Figs.~\refpanels{fig:rdf_perp}{a}{c}  at the same parameters as before. In the passive limit, intermediate- and long-range correlations are observed at $\lambda=6.25$ and $\lambda=12.25$, respectively, highlighting the ordering of columnar clusters. We note that the peak at $\approx 1\sigma$ at $\lambda=12.25$ can be attributed to the pair-bundling of chains [Fig.~\ref{fig:ave_clust_size}]. As activity is increased, spatial correlations are progressively lost, albeit they persist at $f^{a*}$= 20 and $\lambda=12.25$. Therefore, the enhanced branching or shortening of the structures induced by activity is accompanied by a loss of spatial correlations perpendicular to the field.


\bibliography{bib}

\end{document}

%% file: table_steady_states.tex
\begin{table*}
\caption{\label{tab:classification}
Classification of structural states of dipolar ABPs based on the polymerization degree $\phi_p$ [Eq.~\eqref{eq:phi_pol}], the fraction of particles in the largest cluster $\phi_{\text{max}}$ [Eq.~\eqref{eq:phi_max}], and the mean polarization $P$ [Eq.~\eqref{eq:normalized_polarization}].}
\begin{ruledtabular}
\begin{tabular}{lccc}
\textbf{State} & \textbf{Polymerization} ($\phi_p$) & \textbf{Percolation} ($\phi_{\text{max}}$) & \textbf{Polarization} ($P$) \\
\hline
Gas                         & $\phi_p < 0.5$               & $\phi_{\text{max}} < 0.5$            & $P < 0.5$ \\
Polarized Gas               & $\phi_p < 0.5$               & $\phi_{\text{max}} < 0.5$            & $P \geq 0.5$ \\
String Fluid                & $\phi_p \geq 0.5$            & $\phi_{\text{max}} < 0.7$            & $P < 0.5$ \\
Polarized String Fluid      & $\phi_p \geq 0.5$            & $\phi_{\text{max}} < 0.7$            & $P \geq 0.5$ \\
Percolated Network          & $\phi_p \geq 0.5$            & $\phi_{\text{max}} \geq 0.7$         & $P < 0.5$ \\
Polarized Percolated Network& $\phi_p \geq 0.5$            & $\phi_{\text{max}} \geq 0.7$         & $P \geq 0.5$ \\
\end{tabular}
\end{ruledtabular}
\label{tab:orderparameters}
\end{table*}